\documentclass[aps,prl,reprint,twocolumn,notitlepage,floatfix,tightenlines,superscriptaddress]{revtex4-1}
\usepackage[latin9]{inputenc}
\usepackage{natbib}
\makeatletter

\usepackage{amsfonts}
\usepackage{amsmath}
\usepackage{amssymb}
\usepackage[T1]{fontenc}
\usepackage{graphicx}

\numberwithin{equation}{section}

\usepackage[colorlinks=true,citecolor=blue,linkcolor=blue,urlcolor=blue]{hyperref}

\makeatother

\newcommand{\vect}[1] {\mathbf{#1}}   
\newcommand{\tens}[1] {\tensor{#1}}   

\newcommand{\dif} {\mathrm{d}}

\newcommand{\AF} {A_{\text{f}}}
\newcommand{\AB}{A_{\text{b}}}
\newcommand{\measureb} { \frac{\dif^2 p}{(2\pi)^2}\frac{\dif z}{2\pi} }
\newcommand{\measuref} { \frac{\dif^2 k}{(2\pi)^2}\frac{\dif z}{2\pi} }

\newcommand{\LSCO}{ La$_{2-x}$Sr$_x$CuO$_4$ }
\newcommand{\BSCCO}{Bi$_2$Sr$_2$CaCu$_2$O$_{8+\delta}$}
\newcommand{\YBCO}{YBa$_2$Cu$_3$O$_{6+\delta}$}
\newcommand{\HgBCO}{HgBa$_2$CuO$_{4+\delta}$}

\def\b{\mathrm{b}}

\def\pg{\mathrm{pg}}

\def\mf{{\mathrm{mf}}}

\def\GL{\mathrm{GL}}
\def\lb{\ell_{\mathrm{B}}}
\def\H{\mathrm{H}}
\def\R{\mathrm{R}}

\def\pair{\mathrm{pair}}
\def\tr{\mathrm{tr}}
\def\phikp2{\varphi_{\vect{k}-\vect{p}/2}^2}

\def\phik{\varphi_\vect{k}}
\allowdisplaybreaks[4]

\begin{document}

\title{Unified approach to electrical and thermal transport in high-$T_c$ superconductors}

\author{Rufus Boyack}
\affiliation{D\'epartement de physique, Universit\'e de Montr\'eal, Montr\'eal, Qu\'ebec H3C 3J7, Canada}
\author{Zhiqiang Wang}
\affiliation{James Franck Institute, University of Chicago, Chicago, Illinois 60637, USA}
\author{Qijin Chen}
\affiliation{Hefei National Laboratory for Physical Sciences at the Microscale University of Science and Technology of China, Shanghai 201315, China}
\author{K. Levin}
\affiliation{James Franck Institute, University of Chicago, Chicago, Illinois 60637, USA}

\date{\today}
\begin{abstract}
In this paper we present a consolidated equation for all low-field transport coefficients,
based on a reservoir approach developed for non-interacting quasiparticles.
This formalism allows us to treat the two distinct types of charged (fermionic and bosonic)
quasiparticles that can be simultaneously present, as for example in superconductors.
Indeed, in the underdoped cuprate superconductors these two types of carriers
result in two onset temperatures with distinct features in transport: $T^*$, where the fermions first experience an excitation (pseudo)gap,
and $T_c$, where bosonic conduction processes are dominant and often divergent.
This provides the central goal of this paper, which is to address the challenges in thermoelectric transport that stem from having
two characteristic temperatures as well as
two types of charge carriers whose contributions can in some instances enhance each other and in others compete. 
We show how essential features of the 
cuprates (their bad-metal character and the presence of Fermi arcs)
provide an explanation for the classic pseudogap onset signatures at $T^*$ in the longitudinal resistivity, $\rho_{xx}$.
Based on the fits to the temperature-dependent $\rho_{xx}$, we present the
implications for all of the other thermoelectric transport properties.
\end{abstract}

\maketitle

\section{Introduction}

There is renewed interest in thermoelectric transport properties in the condensed matter community, 
motivated in part by the refined experimental capabilities which address the difficult measurements~\cite{Kasahara2018,Banerjee2018,Grissonnanche2019,Li2020,Grissonnanche2020} of the small thermal Hall effect. 
The thermal Hall effect has attracted additional excitement because of topological signatures~\cite{Xiao2006,Kasahara2018b,Banerjee2018} which we now understand can be embedded in these data.
Moreover, with better experimental understanding of
strongly-correlated materials, such as the high-temperature superconductors,
this has called attention to the need to understand transport properties in a holistic way, rather than to focus on any particular quantity alone.
Among the issues of interest~\cite{Obraztsov1964,Cooper1997,Qin2011} are subtle but important features related to magnetization effects, which must be incorporated in a proper treatment of thermoelectric transport properties. 

This leads to the goal of the present paper, which is to address the full complement of transport coefficients 
in the normal-state ($T>T_c$) of strongly-correlated superconductors,
focusing on the weak magnetic field regime.
We assume that this normal state is characterized by a mixture of fermions and fluctuating Cooper pairs. 
However, in contrast to conventional fluctuation theory~\cite{VarlamovBook}, the fermions themselves
are under the influence of pairing fluctuations and this in turn feeds back to renormalize the pairs.
Both components contribute to the thermoelectric transport properties, and when the two are treated as non-interacting quasiparticles we demonstrate, quite remarkably, 
that their contributions can be consolidated into a single equation for all transport coefficients.

Our derivation of the transport coefficients is based on the introduction of a grand-canonical reservoir containing localized, charged particles
(such as proposed by Caldeira and Leggett~\cite{Caldeira1983}) which can be either bosonic or fermionic.
We couple the reservoir to a system (consisting of either non-interacting fermions or bosons) with the same statistics. 
Because the principal system is open, dissipation is automatically generated in the transport expressions after integrating out the reservoir degrees of freedom. 
Furthermore, this approach allows a more direct treatment of temperature gradient perturbations and enables us to avoid the complications of Luttinger's~\cite{Luttinger1964} gravitational potential approach for deriving thermal transport coefficients.
Our single equation for the full set of thermoelectric transport coefficients depends only
on the particle spectral functions.
Importantly, our formulae have properly taken into account the magnetization currents~\cite{Obraztsov1964,Cooper1997,Qin2011} and satisfy the Onsager reciprocal relations. 

In the mixture of fermions and fluctuating Cooper pairs, these latter, bosonic contributions are often described by Gaussian fluctuation theory. 
This paper goes beyond this conventional fluctuation scheme. The framework we use~\cite{Chen2005} 
can be interpreted as an extension of
self-consistent Hartree fluctuation theory~\cite{Ullah1991,Hassing1973}, which produces a pairing-induced pseudogap in the fermionic energy spectrum. We define the onset temperature of this pseudogap as $T^*$, which can be
larger than $T_c$ by orders of magnitude as the pairing attraction becomes progressively stronger.
The formation of the pseudogap reflects the fact that electrons and fluctuating Cooper pairs are strongly intertwined.
Both contributions to transport should be considered simultaneously. 

While in this paper we emphasize both fermionic and bosonic transport contributions, the literature in the cuprates exhibits a kind of dichotomy. 
There has been a focus on theories in which the transport is dominated by bosonic contributions where $T^*$ plays no role. 
These derive from conventional Gaussian fluctuations~\cite{Varlamov1990,Varlamov1992,Ussishkin2002,Niven2002,Ussishkin2003,Serbyn2009}, which are important in the immediate vicinity of $T_c$.
Additionally, there has been a focus on transport associated with pseudogapped fermions~\cite{Storey2013, Storey2016, Verret2017}, 
where $T_c$ plays little or no role. 
For this case, however, a variety of different origins of the pseudogap have been contemplated. 

Here we argue that both the fluctuating Cooper pairs and pseudogapped fermions should be present in strongly-correlated superconductors.
Moreover, if the pseudogap is due to pairing fluctuations, the pairs and fermions are not independent. 
Thus, we find the size of the fermionic gap reflects the bosonic binding energy and constrains the number of preformed pairs. 
Due to the finite lifetime of these pairs, and their $d$-wave pairing nature, the pseudogap leads to a broadening of the nodes in the fermionic energy spectrum, 
which can, in the cuprates, be associated with Fermi arcs~\cite{Chen2008}.

Given that both the fermions and bosons contribute to thermoelectric transport, 
a key question with strong experimental implications is: under what circumstances can the bosonic
transport contributions be visible, as compared to those from the fermions, at relatively high temperatures around $T^*$?
Interestingly, we find that the bad-metal character of the cuprates enables the bosonic terms to dominate their fermionic counterparts, making the former more evident well outside the conventional fluctuation regime.
We illustrate these points using a Fermi arc + preformed Cooper pair model for the cuprates. 
In order to constrain the phenomenological parameters in our model, we fit the theoretical temperature-dependent DC resistivity, $\rho_{xx}(T)$, to that of a typical cuprate with a pseudogap.  
This serves to disentangle the relative weights of the fermionic and bosonic contributions;  we then outline the resulting implications for the entire complement of transport properties.  

As one last argument for our more holistic approach to transport, we note that this enables us to quantitatively evaluate the open-circuit corrections to transport coefficients.
These have recently been of interest~\citep{Kavokin2020} in the context of thermal Hall measurements. 
Here, we quite generally quantify these contributions and find that they are negligibly small.

The remainder of the paper is organized as follows. In Sec.~\ref{sec:Transport} we define transport coefficients in general and discuss
their symmetry properties. In Sec.~\ref{sec:TransportBoson} we present our theoretical 
formalism and results for the bosonic thermoelectric transport.
Section~\ref{sec:GG0} reviews our
strong-pairing fluctuation theory. The inclusion of the fermionic
contributions is discussed in Sec.~\ref{sec:Fermions}, while an overview of the numerical
results appears
in Sec.~\ref{sec:Numerics}.
Section~\ref{sec:opencircuit} contains estimates of the usually neglected open-circuit correction terms
using our numerical approach as well as
experimental data on the cuprates.
We conclude our paper with Sec.~\ref{sec:Conclusions}.
In addition, we make a number of details available in several appendices. Appendix~\ref{app:GL} gives
a summary of transport in conventional
superconducting fluctuation theory; Appendix~\ref{app:GoodBad} presents a comparison of transport in bad and good metals.
Finally,  Appendices~\ref{app:Comparison} and \ref{app:unit} provide detailed comparisons between our numerical results 
(for the physical case of highly resistive or bad metals)
and cuprate experiments.

\section{General aspects of transport theory}

\label{sec:Transport}
\subsection{General transport coefficients}

The transport coefficients are defined within linear-response theory, where the external perturbations consist of an electric field $\vect{E}$ and a temperature gradient $\vect{\Theta}=-\nabla{T}$.
The pertinent equations for the (transport) electric current $\vect{J}^e_{\tr}$ and the (transport) heat current $\vect{J}^h_{\tr}$ are defined in terms of these coefficients by
\begin{equation}\label{eq:transport}\begin{split}
\vect{J}^e_{\tr}&=\tens{\sigma}\vect{E}+\tens{\beta}\vect{\Theta},\\
\vect{J}^h_{\tr}&=\tens{\gamma}\vect{E}+\tens{\kappa}\vect{\Theta}.
\end{split}\end{equation}
The electrical conductivity tensor is denoted by $\tens{\sigma}$, the
thermal conductivity tensor is $\tens{\kappa}$, and the thermoelectric
tensors are $\tens{\beta}$ and $\tens{\gamma}$.  
Solving Eq.~\eqref{eq:transport} for $\vect{E}$ gives
\begin{equation}\label{eq:ElecField}
\vect{E}=\tens{\sigma}^{-1} \left(\vect{J}^e_{\tr}-\tens{\beta} \vect{\Theta}\right).
\end{equation}
As discussed in Refs.~\citep{Langer1962,Luttinger1964,Luttinger1964b,AbrikosovBook}, thermal and thermoelectric response measurements are carried out under open-circuit conditions, 
where the transport electric current is set to zero: $\vect{J}^e_{\tr}=0$. 
In the absence of a particle current, Eq.~\eqref{eq:ElecField} shows that a nonzero temperature gradient produces an electric field. 
This can be understood physically from the fact that a temperature gradient causes a diffusion of particles, which 
then sets up the field. 

The thermopower tensor, denoted by $\tens{S}$, is an important transport quantity, and it is defined~\cite{Ziman,VarlamovBook} by expressing Eq.~\eqref{eq:ElecField} as
$\vect{E}=\tens{\rho}\vect{J}^e_{\tr}-\tens{S}\vect{\Theta}$, where $\tens{\rho}$ is the resistivity tensor. 
Comparing this expression with Eq.~\eqref{eq:ElecField} gives
\begin{align}
\label{eq:Rho1}\tens{\rho}&=\tens{\sigma}^{-1},\\
\label{eq:Thermo1}\tens{S}&=\tens{\sigma}^{-1}\tens{\beta}=\tens{\rho}\tens{\beta}.
\end{align}
In the presence of an external magnetic field $\vect B=B\hat{z}$ (perpendicular to the two-dimensional (2d) system), 
the transverse components of these tensors are nonzero and of considerable interest. The transverse resistivity, $\rho_{yx}$, measured in the absence of a temperature gradient, is related to the Hall coefficient $R_{\H}$ via
$\rho_{yx}=BR_{\H}$. 
This is given in terms of the components of the electrical conductivity by
\begin{equation}
\label{eq:RH1}
R_{\H} = \frac{\rho_{yx}}{B}=\frac{1}{B}\frac{\sigma_{xy}}{\sigma_{xx}^2 + \sigma_{xy}^2}.
\end{equation}
Similarly, the transverse thermopower is related to the Nernst coefficient, $\nu$,  via $S_{yx}=-B\nu$, with
\begin{equation}
\label{eq:Nernst1}
\nu = -\frac{S_{yx}}{B}=\frac{1}{B} \left( \frac{\beta_{xy}\sigma_{xx} - \beta_{xx}
\sigma_{xy}}{\sigma_{xx}^2 + \sigma_{xy}^2} \right).
\end{equation}

Another important transverse transport quantity is the thermal Hall conductivity. 
Setting $\vect{J}_{\tr}^e$ to zero in Eq.~\eqref{eq:ElecField} and then inserting this expression for the electric field into Eq.~\eqref{eq:transport} yields a heat current
\begin{equation}
\label{eq:Jh1}
\vect{J}^{h}_{\tr}=\left[\tens{\kappa}-\tens{\gamma}\tens{\sigma}^{-1}\tens{\beta}\right]\vect{\Theta}\equiv\tens{\widetilde{\kappa}}\vect{\Theta}.
\end{equation}
The last equivalence defines the measured thermal conductivity tensor $\tens{\widetilde{\kappa}}$ via Fourier's law of heat conduction. 
Written out explicitly, the thermal Hall conductivity is
\begin{equation}
\label{eq:kappayx}
\widetilde{\kappa}_{xy}=\kappa_{xy}-T\frac{\left(\beta_{xy}^{2}-\beta_{xx}^{2}\right)\sigma_{xy}+2\beta_{xx}\sigma_{xx}\beta_{xy}}{\sigma_{xx}^{2}+\sigma_{xy}^{2}},
\end{equation}
where use has been made of the Onsager relations between $\tens{\gamma}$ and $\tens{\beta}$. This expression highlights how understanding thermal transport requires knowledge of the magnitude of both the electrical and thermoelectric conductivities. These
interconnections between electrical and thermal response serve to emphasize the value
of a unified theory that deals with all transport coefficients simultaneously.

\subsection{Onsager relations and particle-hole symmetry}
\label{sec:Onsager}

Due to time-reversal symmetry in the underlying equations of motion, the transport coefficient tensors in Eq.~\eqref{eq:transport} obey
the following Onsager reciprocal relations: $\sigma_{ij}(\vect{B})=\sigma_{ji}(-\vect{B})$, $\beta_{ij}(\vect{B})=\gamma_{ji}(-\vect{B})/T$, and $\kappa_{ij}(\vect{B})=\kappa_{ji}(-\vect{B})$.

These coefficients additionally possess particle-hole transformation properties.
The particle-hole symmetry operator is denoted by $\mathcal{C}$. 
Under the action of this operator, $\vect{E}\rightarrow-\vect{E}, \vect{B}\rightarrow-\vect{B},$ and $\vect{J}^{e}_{\tr} \rightarrow-\vect{J}^{e}_{\tr}$, whereas $T$ and $\vect{J}^{h}_{\tr}$ are invariant. 
Using these relations in Eq.~\eqref{eq:transport}, the transformation properties for the transport coefficients are
$\mathcal{C}\sigma\left({\bf B}\right)\mathcal{C}^{-1}=\sigma\left(-{\bf B}\right), \mathcal{C}\beta\left({\bf B}\right)\mathcal{C}^{-1}=-\beta\left(-{\bf B}\right),$ and $\mathcal{C}\kappa\left({\bf B}\right)\mathcal{C}^{-1}=\kappa\left(-{\bf B}\right)$.
From these relations it follows that a particle-hole symmetric system has $\beta_{xx}=\sigma_{xy}=\kappa_{xy}=0$.

\section{Transport coefficients for noninteracting particles}
\label{sec:TransportBoson}

In this section we present a methodology that enables all of the transport coefficients in Eq.~\eqref{eq:transport} to be expressed in terms of a simple formula that depends only on the relevant spectral functions and associated vertex functions. 
This theory applies to free bosonic and fermionic theories, with arbitrary single-band dispersion relations that preserve time reversal, spatial inversion,
and rotation symmetries. 
It is thus applicable to the bosonic contribution in the Ginzburg-Landau (GL) fluctuation theory of superconductors~\cite{VarlamovBook},
as well as to a strong-pairing fluctuation theory~\cite{Chen2005} that incorporates a normal-state cuprate pseudogap,
to be discussed in Sec.~\ref{sec:GG0}. 

For concreteness we first consider the case of non-interacting bosons, presenting
an outline of the derivation and leaving 
a more detailed exposition to a forthcoming paper. This work is based on an approach first introduced in
Ref.~\cite{Tan2004}. 

To derive the coefficients in Eq.~\eqref{eq:transport} we couple our principal system to a Caldeira-Leggett thermal reservoir with
localized particles~\cite{Tan2004}. The reservoir particles have the same type of charge and statistics as those of the principal system. 
The system achieves equilibrium by exchanging particles and energy with the reservoir.
Applying a perturbation, either $\vect{E}$ or $\vect{\Theta}$ in the presence of the magnetic field $\vect B$, one can write down the equation of motion for both the
principal system and the reservoir particles in the presence of the perturbation, and then integrate out the reservoir particles
by absorbing their effects as a phenomenological local self energy into the definition of the Green's function
associated with the principal system. 
Using these perturbed Green's functions we then compute the microscopic current densities, $ \langle \hat{\vect{J}}^e \rangle_{\vect{E}, \vect{\Theta}}$
 and $\langle \hat{\vect{J}}^h \rangle_{\vect{E}, \vect{\Theta}}$, to linear order in $\vect{E}$ or $\vect{\Theta}$ and in $\vect B$, which can be 
written compactly as
\begin{equation} \label{eq:microscopicJ}
\begin{bmatrix}
\langle \hat{ \vect{J}}^e \rangle_{\vect{E}, \vect{\Theta}} \\
\langle \hat{\vect{J}}^h \rangle_{\vect{E}, \vect{\Theta}}
\end{bmatrix}
=
\begin{bmatrix}
\tens{L}_{e,e}    \quad   &    \tens{L}_{e,h} \\
 \tens{L}_{h,e}    \quad  &    \tens{L}_{h,h}
\end{bmatrix}
\begin{bmatrix}
\vect{E} \\
\vect{\Theta}
\end{bmatrix}.
\end{equation}
Here, $\hat{ \vect{J}}^e$ and $\hat{ \vect{J}}^h$ are the electric and heat current density operators; 
$\tens{L}_{e,e}, \tens{L}_{e,h}, \tens{L}_{h,e}$ and $\tens{L}_{h,h}$ are four tensorial coefficients,
which can be fully expressed in terms of the particle's Green's functions, retarded and/or advanced.  

It is well known that the microscopic currents in Eq.~\eqref{eq:microscopicJ} are a subset of
the macroscopic transport currents in Eq.~\eqref{eq:transport}~\cite{Cooper1997,Qin2011}. 
To obtain the transport currents one needs to subtract the divergence-free currents due to the charge magnetization $\vect{M}^{e}$ and the heat magnetization $\vect{M}^{h}$. 
As elaborated in Refs.~\cite{Cooper1997,Qin2011}, in the presence of $\vect{E}$ or $\vect{\Theta}$ perturbations, the correct subtractions are
\begin{equation}\label{eq:Currents}
\begin{bmatrix}
\vect{J}_{\tr}^e \\
\vect{J}_{\tr}^h
\end{bmatrix}
=
\begin{bmatrix}
\tens{L}_{e,e}  \quad  & \tens{L}_{e,h} - \frac{\partial \vect{M}^e }{\partial T} \times       \\
 \tens{L}_{h,e} - \vect{M}^e \times    \quad  &  \tens{L}_{h,h} - \frac{\partial \vect{M}^h }{\partial T} \times
\end{bmatrix}
\begin{bmatrix}
\vect{E}    \\
\vect{\Theta}
\end{bmatrix},
\end{equation}
The transport coefficients are then defined by comparing this equation with Eq.~\eqref{eq:transport}. 
The magnetization currents for the thermoelectric tensors $\tens{\beta}$ and $\tens{\gamma}$, the off-diagonal 
magnetization terms in Eq.~\eqref{eq:Currents}, were first derived by Obraztsov~\cite{Obraztsov1964}.
Including these magnetization currents is vital to ensure the Onsager reciprocal relations and the laws of thermodynamics are obeyed~\cite{Cooper1997,Qin2011}. 

In our reservoir approach, we express both $\tens{L}_{ij}$ and $\vect{M}^i$, with $\{i,j\} \in \{ e, h\}$, in terms of Green's functions.
Although each of the two generally involves complicated combinations of retarded and advanced Green's functions, 
the final answers for the transport coefficients on the right-hand side of Eq.~\eqref{eq:Currents} turn out to be simply
written in
terms of spectral functions and current vertices. We find that the matrix appearing in Eq.~\eqref{eq:Currents}
can be consolidated~\cite{Tan2004} into the form: 
\begin{equation}
J^n_{\text{(tr)}a}=\sum_{n'=0}^1\sum_b\mathcal{L}^{nn'}_{ab}E^{n'}_b,
\end{equation}
with
\begin{align} \label{eq:Lresp}
\mathcal{L}^{nn'}_{ab}=\int\measureb z^{n+n'}\biggl[\frac{q^{2-n-n'}}{2T^{n'}}v_{a}v_{b} \AB^2(z,\vect p)\nonumber\\
+\frac{q^{3-n-n'}}{6T^{n'}} B \epsilon_{cd}  v_{a}v_{c}v_{bd}   \AB^3(z,\vect p)\biggr] b^{(1)}(z).
\end{align}
Here, $\AB(z,\vect p)$ is the bosonic spectral function.
To arrive at this compact form we have introduced the notation  $E^0_a\equiv E_a$ and $E^1_a\equiv-\partial_a T$,
where $a=\{x,y\}$. 
Here $b^{(1)}(z)\equiv -\partial b(z)/\partial z$ where  $b(z)=\left(e^{z/T}-1\right)^{-1}$ is the Bose-Einstein distribution function (we set $\hbar=k_B=1$ and will restore these units only when
necessary).
The relations between $\mathcal{L}^{nn'}_{ab}$ and the coefficients in Eq.~\eqref{eq:transport} are: 
$\sigma_{ab}=\mathcal{L}^{00}_{ab}$, $\beta_{ab}=\mathcal{L}^{01}_{ab}$, $\gamma_{ab}=\mathcal{L}^{10}_{ab}$ and $\kappa_{ab}=\mathcal{L}^{11}_{ab}$. 

In Eq.~\eqref{eq:Lresp}, $z$ is the real frequency and $q$ is the charge of the particles in the
system:  $q=-e< 0$ for electrons and $q=-2e$ for fluctuating Cooper pairs.
We define $v_a$ as the $a$-component velocity whose wavevector $\vect{p}$ dependence has been suppressed for brevity
while $v_{ab}$ is the $ab$-component of the inverse effective-mass tensor, which is also $\vect{p}$ dependent. 
In the second term of Eq.~\eqref{eq:Lresp}, $\epsilon_{cd}$ is the 2d Levi-Civita symbol. 
Because of the underlying spatial inversion and rotation symmetries of the band dispersion, the first term in Eq.~\eqref{eq:Lresp} 
is longitudinal, while the second term is transverse. 

A few comments are in order concerning Eq.~\eqref{eq:Lresp}, as
it might seem rather unexpected that there exists a single, closed-form, Kubo-like expression from which
all of the transverse and longitudinal transport coefficients can be obtained. 
Notably, for the transverse contribution,
this result also includes magnetization corrections in addition to the intrinsic $\tens L_{ij}$ terms.
That this is possible partially follows from confining our attention to the weak magnetic field limit.

We note that from Eq.~\eqref{eq:Lresp} it is easy to verify consistency with the Onsager reciprocal relations, i.~e., 
$T^{n'}\mathcal{L}^{nn'}_{ab}(\vect{B})=T^{n}\mathcal{L}^{n'n}_{ba}(-\vect{B})$ (no summation).
We emphasize that this result is a consequence of including the magnetization terms in Eq.~\eqref{eq:Currents}. 

Next we briefly compare our approach to other methods in the literature. 
In a classic series of papers, Luttinger~\cite{Luttinger1964,Luttinger1964b} developed an approach which
requires introducing a source $\psi$ for perturbations in the energy density,
in analogy with the way the vector potential $\vect{A}$ is used to initiate changes in the electric current.
The field $\psi$ acts as a source that enables describing a local temperature $T(\vect{r})$, which is distinct from the equilibrium temperature.
This formalism builds on the Einstein relation which asserts that coefficients of gradients in $\psi$ and $T$ must be equal;
this is analogous to the other well-known Einstein relation that coefficients of gradients in chemical potential $\mu$ and electric potential $\phi$ are the same for electrical response~\cite{Ziman,Cooper1997}.
The Luttinger approach was implemented in more detail by Cooper et al.~\cite{Cooper1997}, 
who derived the magnetization current contributions  for all of the transport coefficients.

While our formalism is different, we equivalently include these same magnetization current effects here.
We emphasize that in our approach to electrical and thermal transport the thermal response is naturally deduced from temperature fluctuations about the equilibrium temperature set by the reservoir.
The advantages of the heat-reservoir approach are that it provides a direct method to derive thermal transport,
avoiding the more abstract and technically difficult Luttinger formalism.
In this way, we formulate the theory with $\nabla\psi=0$ from the outset.

\section{Transport due to strong-pairing fluctuations: bosons with pseudogap effects}
\label{sec:GG0}

We apply the general expression for transport coefficients in Eq.~\eqref{eq:Lresp} to a
model that consists of a mixture of fluctuating Cooper pairs and electrons, in order to address transport
properties of the pseudogapped normal state of cuprates~\cite{Timusk1999}. 
In reality the fluctuating pairs and electrons are constantly interconverting.
Importantly, the effects of this interconversion can be treated in a self-consistent, mean-field manner,
where they are manifested as a pseudogap in the electron energy spectrum. This excitation gap, 
at the same time, feeds back to alter the general physical properties of the Cooper pairs.

\subsection{Preformed pairs and the pseudogap}

The observations that the pseudogap is associated with
a reduction in carrier 
number~\citep{Taillefer2019} have led many to argue against
the concept of preformed pairs as the origin of the pseudogap.
Indeed, the notion that the pseudogap arises in this way has gone in and out of favor with time.
Indications for some unspecified form of additional order which onsets at $T^*$, as well as alternative experiments, 
have more recently been cited as evidence against a preformed-pair theory.
Among these experiments are (i) evidence for two-gap physics~\cite{Lee2007} in which the nodal and anti-nodal gaps have different temperature dependences, and
(ii) the existence of Fermi arcs was also argued to be difficult to understand~\cite{LeeSenthil2020}, at least within one preformed-pair framework.
Finally, (iii) there are claims of particle-hole asymmetry in the fermionic quasiparticle energy dispersion in the pseudogap
phase~\cite{Hashimoto2010}, although this is not
substantiated by other experiments~\cite{Kanigel2008,BCSnote}.

In this paper we explore the field-dependent transport in a preformed-pair scenario.
Here we are motivated by the anomalously high transition temperatures which support the view that the cuprates are in
an intermediate state between BCS and Bose-Einstein condensation (BEC)~\cite{Legget1982}. 
Importantly, there is well supported evidence
that in the strong-pairing preformed-pair theory the observations (i) and (ii) are in fact fully consistent. 
For point (i) a number of references address the two gap dichotomy \cite{Chien2009,Wulin2010},
while the presence of Fermi arcs is addressed in this paper. 
We emphasize that there are multiple flavors of preformed-pair scenarios; 
the one we consider here has a laboratory realization in ultracold Fermi gas superfluids~\cite{Chen2005}. 
An alternative scenario is the phase-fluctuation picture of Emery and Kivelson~\cite{Emery1995}.
At the very least, the  strong-pairing approach that we adopt represents
a rather benign extension of BCS theory, which appears warranted due to the short coherence length. 

We emphasize that the physical picture introduced in this paper can also be
seen as a natural generalization of self-consistent-Hartree approaches~\cite{Ullah1991,Hassing1973}
to the time-dependent Ginzburg-Landau framework~\cite{Stajic2003}. 
These have been independently advocated~\cite{Ussishkin2002} for addressing cuprate transport.

\subsection{Theoretical approach to strong-pairing fluctuations}
\label{sec:framework}

Our approach to preformed-pair theory is, in some sense, a generalization of conventional superconducting 
fluctuation theory (see Appendix~\ref{app:GL}).
The latter, however, does not explicitly incorporate pseudogap effects, which we argue are associated with 
stronger-than-BCS attractive interactions.
One expects, as the pairing strength increases, the inverse fluctuation propagator transitions from a predominantly 
diffusive to a propagating form, which we write as
\begin{equation}
\label{eq:GG0Prop}
t^{-1}_{\R}\left(z,\vect{p} \right)=Z\left[\kappa z- \vect{p}^{2}/\left(2M_{\text{pair}}\right) - |\mu_{\text{pair}}| +i\Gamma z\right],
\end{equation}
where $\{ Z, \kappa, \Gamma\}$ are all real. 
This expression breaks particle-hole symmetry due to the presence of the $\kappa z$ term.
We will find that $\kappa$, which is equal to $\pm 1$ here, plays an important role in the
transport coefficients $\beta_{xx}, \sigma_{xy}$, and $\kappa_{xy}$.
In conventional fluctuation theories the counterpart to $\kappa$ is small.

We now give a brief overview of the formalism that we use to arrive at the 
pair propagator (or $t$-matrix), $t_{\R}$ in Eq.~\eqref{eq:GG0Prop}.
This is based on a BCS-like structure, extended to include stronger attractive interactions.
We are motivated by the observation that, in BCS theory, the gap equation (here considered for $d$-wave pairing symmetry), 
\begin{equation} \label{eq:BCS}
 0=\frac{1}{g}+\sum_{\vect k} \frac{1 - 2f(E_{\vect k})}{2 E_{\vect k}}\varphi_{\vect{k}}^2, 
\end{equation}
can be understood as a Thouless criterion reflecting a $q = 0$ divergence of a dressed $t$-matrix~\cite{Chen2005}, 
with inverse 
 \begin{equation}
\label{eq:tmatGG0}
t^{-1}(p)=\sum_{k}G(k)G_{0}(p-k)\phikp2+g^{-1}.
\end{equation}
Here $f(x)$ is the Fermi-Dirac distribution function.

In Eq.~\eqref{eq:BCS}, $g<0$ is the strength of the attractive $d$-wave pairing interaction $V_{\vect{k},\vect{k}^\prime} = g\phik\varphi_\vect{k'}$, where $\varphi_{\vect{k}}= \cos k_x  -\cos k_y$ is the $d$-wave pairing form factor;
$ E_{\vect k} = \sqrt{\xi_{\vect k}^2 + \Delta_{\mf}^2\varphi_{\vect{k}}^2}$ and
the underlying bare fermion dispersion is $\xi_{\vect k} = \epsilon_{\vect k} -\mu_F^{}$,
where $\mu_F^{}$ is the fermionic chemical potential. 
Here, $\Delta_{\mf}$ is the temperature-dependent BCS mean-field (mf) gap.
In Eq.~\eqref{eq:tmatGG0}, $G_0(k)=\left(i\omega_n - \xi_{\vect k}\right)^{-1}$ and  $G(k) \equiv \left[ G_0^{-1}(k) - \Sigma(k) \right]^{-1}$ are the bare and dressed fermionic Green's functions, respectively,
where $\Sigma(k)=- (\Delta_{\mf} \varphi_{\vect k} )^2 G_0(-k)$ is the superconducting self energy. 
We define $k=(i\omega_n, \vect k)$ and $p=(i \Omega_m, \vect{p})$ as two four-vectors with $\omega_n=(2n+1)\pi T$ and $\Omega_m=2 m \pi T$.

The $t$-matrix in Eq.~\eqref{eq:tmatGG0} involves one bare and one dressed Green's function, as has been recognized in the literature~\cite{Kadanoff1961,Patton1971}.
If one expands $t^{-1}(p)$ at $p=0$ in a Taylor expansion and analytically continues the result to real frequencies, as in Eq.~\eqref{eq:GG0Prop}, one finds that the Thouless criterion, $t^{-1}(p=0)=0$, can be regarded as a 
BEC condition for the pair chemical potential: $\mu_{\pair}=0$ for all $T \leq T_c$. 

We note that this BEC condition describes the approach to condensation from above $T_c$. 
In this way, one should view the associated $t$-matrix in the Thouless criterion as characterizing the non-condensed pairs.
To extend this approach to strong pairing, we use the same $t$-matrix as in Eq.~\eqref{eq:tmatGG0} but with a crucial difference from strict BCS theory 
in the fermion self energy $\Sigma(k)$. In general $t$-matrix theories, and in ours in particular,
\begin{equation}
\label{eq:GG0SelfEnergy}
\Sigma(k)=\sum_{p}t(p)G_{0}(p-k)\phikp2 .
\end{equation}

Here we represent the non-condensed pairs through contributions from the $q\ne 0$ component of $t(q)$.
By numerically solving the coupled Eqs.~\eqref{eq:tmatGG0} and \eqref{eq:GG0SelfEnergy}, 
one can determine $t(q)$. This is, however, very challenging.  

For this reason, we adopt the so-called pseudogap (pg) approximation, 
in which one observes that $t(p)$ is strongly peaked about $p=0$
so that one can approximate $t(p)$ in Eq.~\eqref{eq:tmatGG0} (after analytical continuation)  by the form of $t_{\R}(z,\vect{p})$ in Eq.~\eqref{eq:GG0Prop}.
Simultaneously, the approximate self energy due to the non-condensed pairs 
is given by $\Sigma(k) \approx - (\Delta_{\pg} \phik)^2 G_0(-k)$, with
\begin{equation}
\label{eq:Deltapg}
\Delta_{\pg}^2 (T) = -\sum_{p \ne 0}t(p), \quad T \lesssim T_{c}.
\end{equation}
Note the $p=0$ component of $t(p)$ (which corresponds to the condensate)
is necessarily excluded in the above summation.
The above approximation is valid for $|\mu_{\pair}|$ very small, that is, below or only slightly above $T_c$.
Precisely at $T=T_{c}$, all pairs are non-condensed and thus the condition $\mu_{\pair}=0$ requires~\cite{Chen2005} that $\Delta_{\pg}(T_c) = \Delta_{\mf}(T_c)$.

The focus of this paper is, however, on transport in the normal state
over the entire temperature range between $T_c$ and $T^*$. Indeed,
one observes that $T^* \gg T_c$ when the pairing is strong. 
In contrast to the above treatment for $T\lesssim T_c$, which has a rather precise microscopic basis~\cite{Chen2005},
we must make some simple, but physical assumptions to extend it well away from $T_c$.
We assume that in the normal state
\begin{equation}
\Delta_{\pg}(T) = \Delta_{\mf}(T), \quad T \ge T_c.
\end{equation}
As a consequence, $\Delta_{\pg}$ vanishes at the mean-field transition temperature, which we associate with $T^*$,
although, strictly speaking, there should be a gentle crossover to zero here.

We solve Eq.~\eqref{eq:BCS} together with the electron density constraint equation, 
\begin{align}
n_e = \sum_{\vect{k}} \left[1- \frac{\xi_{\vect{k}}}{E_{\vect{k}}} \tanh\left(\frac{E_{\vect{k}}}{2 T}\right)\right], 
\end{align}
to determine $\Delta_{\mf}(T)$ and the fermionic chemical potential $\mu_F(T)$ for given $\{g, T, n_e\}$. 
The coupling constant $g$ is chosen to give the desired $T^*$. 
With $\Delta_{\mf}$ for the excitation gap determined, we can then derive~\cite{QijinThesis} $M_{\pair}$ and $Z$ from the strong-pairing $t$-matrix in Eq.~\eqref{eq:tmatGG0}.

This group of assumptions is rather benign, as it guarantees continuity between the normal-state results and those at $T_c$.
However, understanding $\mu_{\pair}$ in the normal state is more subtle.
There is no reliable way of computing $\mu_{\pair}$ for the whole temperature range $T_c<T<T^*$ in a self-consistent manner.
Here, we take a more  phenomenological approach by interpolating results of $\mu_{\pair}$ in the two limits $T\rightarrow T_c$ and $T\rightarrow T^*$.
For $T$ slightly above $T_c$, we presume that $\mu_{\pair}$ is well described by the conventional fluctuation behavior,
namely, $\mu_{\pair} \approx  (8/\pi)(T_c-T)$~\cite{VarlamovBook}.
This will be justified later in the paper through a comparison of the calculated
longitudinal resistivity with corresponding experiments.
In the other limit, $T\rightarrow T^*$, the number of pairs vanishes, which requires that $\mu_{\pair} \rightarrow -\infty $.
A formula for $\mu_{\pair}$ that satisfies these requirements and that smoothly interpolates between the two limits is~
\footnote{
The logarithmic divergence of $\mu_{\pair}$ at $T^*$ can be motivated as follows.
Assuming that Eq.~\eqref{eq:Deltapg} is applicable for $T>T_c$, Eq.~\eqref{eq:tmatGG0} implies for 2d
$\Delta_{\pg}^2 Z = - M_{\pair} T / (2\pi ) \ln(1-e^{ \mu_{\pair} /T })$, leading to $\mu_{\pair} \propto \ln (T^*-T)$,
where we have assumed that $Z$ and $M_{\pair}$ are finite at $T\approx T^*$ and used $\Delta_{\pg}^2=\Delta_{\mf}^2 \propto (T^*-T)$.
Despite being well motivated, in reality the divergence of $\mu_{\pair}$ at $T=T^*$ may be much less singular than logarithmic.
}.
\begin{equation}
\mu_{\pair}   =  \frac{8}{\pi} (T^*-T_c) \ln\frac{T^* -T}{T^*-T_c}.
\label{eq:intrplmub}
\end{equation}
We emphasize that our results are not in detail sensitive to this particular interpolation.

\subsection{Transport coefficients in the small $\left|\mu_{\pair}\right|$ limit}
\label{sec:divergence}

The central quantity for calculating the bosonic contribution to transport coefficients is
the bosonic spectral function $\AB$ which appears in Eq.~\eqref{eq:Lresp}. 
This is defined in terms of the retarded pairing fluctuation propagator $t_{\R}$ as
~\footnote{The dimension of $\AB$ defined in Eq.~\eqref{eq:BoseSpectral} is different from that of a usual spectral function due to the factor $Z$
in the definition of $t_{\R}$. $Z$ carries the dimension of density of states over energy (see Eq.~\eqref{eq:Zmup}). However, it
drops out in the calculation of the transport coefficients, due to the Ward identity (see the main text).}
\begin{equation}
\AB(z,\vect{p})=\text{Re}[2it_{\R}(z,\vect{p})].
\label{eq:BoseSpectral}
\end{equation}
The velocity components in Eq.~\eqref{eq:Lresp}
are defined by $v_a(\vect{p})\equiv - \partial t_{\R}^{-1}(z,\vect{p})/\partial p_a = Z p_a/M_{\pair}$,
which constitutes a Ward identity~\cite{Ryderbook,Boyack2018}; similarly the inverse effective-mass
component $v_{ab}(\vect{p})\equiv - \partial^2 t_{\R}^{-1}(z,\vect{p})/\left(\partial p_a\partial p_b\right)=Z/M_{\pair} \delta_{ab}$, where $\delta_{ab}$
is the Kronecker delta function. 
With the bosonic parameters $\{M_{\pair}, \mu_{\pair},Z\}$ determined from the previous section, one can readily evaluate all of the bosonic transport coefficients from Eq.~\eqref{eq:Lresp}.
Interestingly, the prefactor $Z$ drops out of the transport coefficients due to a cancellation effect.

For a generic temperature, one needs to resort to numerics to evaluate Eq.~\eqref{eq:Lresp}. 
However, in the limit of $T \rightarrow T_c$ and $\left|\mu_{\pair}\right|/T_{c}\ll1$, when the bosons become critical,
an asymptotic formula for the singular contribution of each bosonic transport coefficient can be analytically deduced. 
Here we insert Eqs.~\eqref{eq:GG0Prop} and \eqref{eq:BoseSpectral} into Eq.~\eqref{eq:Lresp} and then expand the Bose-Einstein distribution function
derivative for small frequencies. 
This enables the frequency integral, and subsequently the momentum integral, to be performed analytically. 

The final results (for 2d) are:
\begin{subequations} \label{eq:divergence}
\begin{align}
\label{eq:Sigmaxx}\sigma_{xx}&=\frac{1}{8\pi}\frac{\left(\kappa^2+\Gamma^{2}\right)}{\Gamma}\frac{q^2}{\hbar}\frac{k_{B}T_c}{|\mu_{\pair}|},\\
\beta_{xx}&=\frac{1}{4\pi}\frac{\kappa}{\Gamma}\frac{qk_{B}}{\hbar}\ln\left|\frac{E_{\Lambda}}{\mu_{\pair}}\right|, \\
\label{eq:Kappaxy}\kappa_{xx}&= \mathrm{constant}, \\
\sigma_{xy}&=\frac{1}{24\pi}\frac{\kappa\left(\kappa^{2}+\Gamma^{2}\right)}{\Gamma^{2}} \frac{q}{e} \frac{q^2}{\hbar}\frac{\xi^{2}_{0}}{\lb^2}\frac{\left(k_{B}T_{c}\right)^2}{\left|\mu_{\pair}\right|^{2}}, \\
\beta_{xy}&=\frac{1}{24\pi}\frac{\left(3\kappa^{2}+\Gamma^{2}\right)}{\Gamma^{2}} \frac{q}{e}\frac{q k_B}{\hbar}\frac{\xi_0^2}{\lb^2} \frac{k_{B}T_c}{|\mu_{\pair}|},\\
\kappa_{xy}&=\frac{1}{4\pi}\frac{\kappa}{\Gamma^2}\frac{q}{e}\frac{k_{B}^2 T_c}{\hbar}  \frac{\xi^{2}_{0}}{\lb^2}\ln\left|\frac{E_{\Lambda}}{\mu_{\pair}}\right|,
\end{align}
\end{subequations}
where we have restored the constants $\hbar$ and $k_{B}$ to make the units explicit.
In these equations $\lb =\sqrt{\hbar/eB}$ ($eB>0$) is the magnetic length,  $E_{\Lambda}$ is an energy cutoff~\cite{Ullah1991,Ussishkin2002}, and
$\xi_{0} \equiv\hbar/ \sqrt{2M_{\pair} k_{B}T_{c} }$.

\section{Contributions from the pseudogapped fermions}
\label{sec:Fermions}

In addition to the preformed Cooper pairs, the (pseudo)gapped electrons are also carriers of charge and energy in the normal state of a strongly-correlated superconductor.
That there are generically two types of contributions to transport is evident from conventional fluctuation theory~\cite{VarlamovBook}, 
where gauge invariance requires that both the Aslamazov-Larkin (bosonic) and Maki-Thompson (MT) plus
density of states (DOS) (fermionic) diagrams must be present.  
In contrast to the conventional MT and DOS contributions, here we emphasize that the normal-state fermions must also acquire an excitation gap.

It is of interest to note that there are several phenomenologically motivated papers which presume two types of carriers. 
Geshkenbein et al.~\cite{Geshkenbein1997} considered non-dispersing bosons of charge $-2e$ 
arising from electrons in the small anti-nodal regime of the Fermi surface
where the fermion dispersion is flat; they also included gapless fermions confined to an extended region around the nodes.
This scenario is different from the present physical picture where the fermions and pairs are continuously inter-converting as in a chemical equilibrium process.
More recently, Lee and collaborators~\cite{LeeSenthil2020} have proposed a model for the high-field case, similar to 
that in Ref.~\cite{Geshkenbein1997} with localized bosons at the anti-nodes.

To account for the fermionic contribution to transport, we use an additional
consolidated transport coefficient equation similar to Eq.~\eqref{eq:Lresp} but for non-interacting fermions:
\begin{align} 
\label{eq:Lresp2}
\mathcal{L}^{nn'}_{ab}= 2 \int\measuref z^{n+n'}\biggl[\frac{q^{2-n-n'}}{2T^{n'}}v_{a}v_{b}\AF^2(z,\vect k) \nonumber\\
+\frac{q^{3-n-n'}}{6T^{n'}} B \epsilon_{cd} v_{a}v_{c}v_{bd}\AF^3(z,\vect k)\biggr]f^{(1)}(z).
\end{align}
This is essentially the same as in
Eq.~\eqref{eq:Lresp} except
for the prefactor of $2$ from spin degeneracy.
Here we define $f^{(1)}(z)= -\partial f(z)/\partial z$ as the Fermi-Dirac function derivative and
$\AF(z,\vect k)$ as the fermionic spectral function which can be written in terms of  the retarded single-particle fermionic Green's function.
The latter is given by
\begin{equation}
\AF(z,\vect k)=\text{Re}[2iG_{\R}(z,\vect k)].
\label{eq:FermiSpectral}
\end{equation}
Here the Green's function associated with the pseudogapped fermions is 
rather generally taken in the literature~\cite{Norman2007} to be of the form:
\begin{equation}
G^{-1}_{\R}(z,\vect k) = z - \xi_{\vect k} - \frac{\Delta_{\pg}^2(T) \varphi_{\vect{k}}^2}{z + \xi_{\vect k} + i \gamma}
+i\Sigma_0(T),
\label{eq:FermiArcG}
\end{equation}
where $\xi_{\vect{k}}$ is the non-interacting electron dispersion.  

Important in Eq.~\eqref{eq:FermiArcG} is that in addition to a broadened BCS-like self energy associated with the excitation pseudogap $\Delta_{\pg}$,
we also include a term $\Sigma_0(T)$ which accounts for the finite lifetime of fermions in the absence of superconducting fluctuations. 
In our calculation, $\Sigma_0(T)$ is a phenomenological function which is assumed to
be linear in $T$. While its physical origin is under debate, this accounts 
for the underlying and universally observed linear-in-temperature resistivity in optimal and underdoped
cuprates at $T>T^*$.
We presume this same term is present both above and below $T^*$.

At a more microscopic level, in our thermal-reservoir approach leading to
Eq.~\eqref{eq:Lresp2}, $\Sigma_0$ can be viewed as a self energy effect arising from integrating out local reservoir degrees of freedom.
This is somewhat problematic for arriving at
the pseudogap self energy term 
in Eq.~\eqref{eq:FermiArcG}
which has a microscopic~\cite{Chen2005} and
phenomenological basis~\cite{Norman2007}. Strictly speaking,
the pseudogap self energy cannot be interpreted in this way because it is $\vect k$ dependent and non-local.
Nevertheless, here we apply
Eq.~\eqref{eq:Lresp2} under the assumption that the pseudogapped fermions can be viewed
as independent quasi-particles and
define $v_a \equiv \partial \xi_{\vect k}/\partial k_a$ and $v_{ab} \equiv \partial^2  \xi_{\vect k}/\left(\partial k_a \partial k_b\right)$ in Eq.~\eqref{eq:Lresp2} using the bare fermion dispersion. This ignores
vertex corrections which have been addressed in previous work~\cite{Scherpelz2014} and found to be reasonably unimportant.
We should note that the vertex corrections in the fermionic transport coefficients were also considered in Ref.~\cite{Kontani2003} in the limit of small fermion scattering rate.
(Magnetization corrections, which are crucial in the present paper, however, were not treated, since they were not pertinent in the limit considered.)

A non-zero value of $\gamma$ in
Eq.~\eqref{eq:FermiArcG}
leads to a broadening of the $d$-wave nodes in the electron energy dispersion. This can be associated with Fermi arcs, which
are an important feature of 
angle-resolved photoemission spectroscopy (ARPES) experiments and associated with the pseudogap
of underdoped cuprates~\cite{Norman2007,Chen2005}.
Its origin at the microscopic level here  is related to the finite lifetime of the non-condensed pairs~\cite{Chen2005},
which is in stark contrast to the infinite lifetime of condensed pairs at $T<T_c$.
While the implications of Fermi arcs for transport properties have been considered before~\cite{Wulin2011,Levchenko2010},
these have been restricted to the case where there are no bosons.

We note that in describing the pseudogap state  many other fermionic models have been proposed~\cite{Norman2007}. 
Notable is the so-called Yang, Rice, Zhang (YRZ)~\cite{Rice2011} model, which
contains a rather similar form for the fermionic Green's function~\cite{Scherpelz2014} in Eq.~\eqref{eq:FermiArcG}
but with some important differences. 
The pseudogap parameter in YRZ is not evidently related to pairing fluctuations and
instead of $\xi_{\vect k}$ in the denominator of the pseudogap self energy term in Eq.~\eqref{eq:FermiArcG} the YRZ model 
introduces a different dispersive parameter ($\epsilon_{\vect k}^{\text{NN}}$)
which contains only nearest-neighbor (NN) hopping without a broadening parameter such as
$\gamma$. Technically this is associated with Fermi pockets rather than Fermi arcs as in Eq.~\eqref{eq:FermiArcG},
and related transport theory has focused on zero temperature~\cite{Storey2013,Storey2016,Verret2017}
in the absence of bosonic degrees of freedom.

It is useful to add here that we are not concerned in this paper with charge-density wave (CDW) effects in the
cuprates. The 
central focus of our cuprate transport analysis is the pseudogap regime. The CDW regime is in rather limited regions of the phase diagram.
It typically spans a limited doping range around the commensurate 1/8 doping; additionally it may be induced by very high magnetic fields.
As observed in Ref.~\cite{Badoux2016},  the pseudogap and charge order are separate phenomena.
Thus, our argument for focusing on the pseudogap and pairing effects is that they are an even more robust feature
of the cuprates than the CDW-ordered phases.

\section{Numerical results}
\label{sec:Numerics}
\subsection{Overview}
 
We have emphasized that in strongly-correlated superconductors it is important to study both bosonic and fermionic contributions to transport. 
The bosonic terms will tend to have a singular structure near $T_c$ and even possibly contribute to
transport at temperatures as high as $T^*$.  Very near $T_c$, this singular structure is similar to
the predictions from GL fluctuation theory.

Our approach to transport builds on rather general assumptions related to claims in the experimental literature. 
The effect of the pseudogap onset on the fermions is to lead to a
decrease in effective fermionic carrier density; 
this corresponds to a tendency towards an \textit{upturn} in $\rho_{xx}$
which sets in with decreasing temperature, $T$,  below $T^*$. 
This upturn is inferred from 
the behavior of the resistivity $\rho_{xx}(T)$ in cuprates when the
superconductivity is suppressed by three different mechanisms --
high magnetic fields, $B$~\cite{Taillefer2019, Ono2000}, extreme underdoping~\cite{Ando2004},
and the destruction of superconductivity due to Zn impurities~\cite{Momono1994,Naqib2005}.

These experiments are relevant to the behavior of the fermionic channel in the superconducting
cuprates due to the widely held belief that such experiments, in particular those at high
fields, reveal the pristine character of the pseudogap.
Importantly, though, what is observed as the signature of the cuprate pseudogap in $\rho_{xx}(T)$ 
for the normal phase is
a \textit{downturn} which sets in at $T^*$ with decreasing $T$.
This could be understood if  [as in 
Eq.~\eqref{eq:intrplmub}], due to stronger pairing attraction, the bosons
are present and contribute to transport away from the critical regime.

In this section we show the above finding, concerning a form of high temperature bosonic paraconductivity, is a consequence of the
fact that cuprates possess two important
features: they are bad metals, and, importantly, also their $d$-wave pseudogap is associated not with gap nodes but with Fermi arcs. 
The first of these implies that the fermions play a dominant role in
the resistivity or, equivalently, a more minimal role in the conductivity.
As a result of the Fermi arcs, the fermionic tendency to enhance the resistivity starting at $T^*$ is  suppressed. 

Our calculations depend on several phenomenological parameters, including $\{\gamma,\Sigma_0\}$  in Eq.~\eqref{eq:FermiArcG} and $\Gamma$ in Eq.~\eqref{eq:GG0Prop}.
The broadening parameter $\gamma$, in part, sets the length of the Fermi arc in Eq.~\eqref{eq:FermiArcG} (the Fermi arc length also depends on $\Sigma_0$, but the dependence is weak). 
As is compatible with ARPES measurements, we consider $\gamma=\widetilde{\gamma}  T$~\cite{Norman1998,Kanigel2006,Norman2007,Chen2008},
with the dimensionless parameter $\widetilde{\gamma}=3.2$.
We determine the remaining two parameters, $\Sigma_0$ and $\Gamma$, by fitting $\rho_{xx}$ to the corresponding
experimental data of a typical cuprate with a pseudogap. Using those constrained parameters, we then calculate all of the other transport coefficients. 

For the fermionic calculation we use $\xi_{\vect{k}} = 2 t [ 2- \cos k_x -\cos k_y] -4 t^\prime [1- \cos k_x \cos k_y] -\mu_F$ for the dispersion in Eq.~\eqref{eq:FermiArcG} 
 with $t^\prime/t=0.35$. Here $\mu_F$ is adjusted to give an electron density $n_e= 0.85$ per CuO$_2$ square, i.~e., hole doping $p=1-n_e=15\%$.
 For definiteness we choose $t=75 \mathrm{meV}$.

\subsection{Longitudinal resistivity $\rho_{xx}$}

The resistivity of underdoped cuprates typically exhibits a linear-in-$T$ behavior at high temperature $T>T^*$,
deriving from a presently unknown mechanism. 
As temperature is decreased there comes a point ($T^*$) when the resistivity exhibits a slight
downturn from linearity which is maintained until essentially at the transition temperature $T_c$ where it
plummets to zero. As is conventional, we take this downturn onset to be a signature of the entrance into
the pseudogap phase. 

\begin{figure*}
\includegraphics[width=.725\textwidth,trim=0mm 10mm 0mm 10mm]
{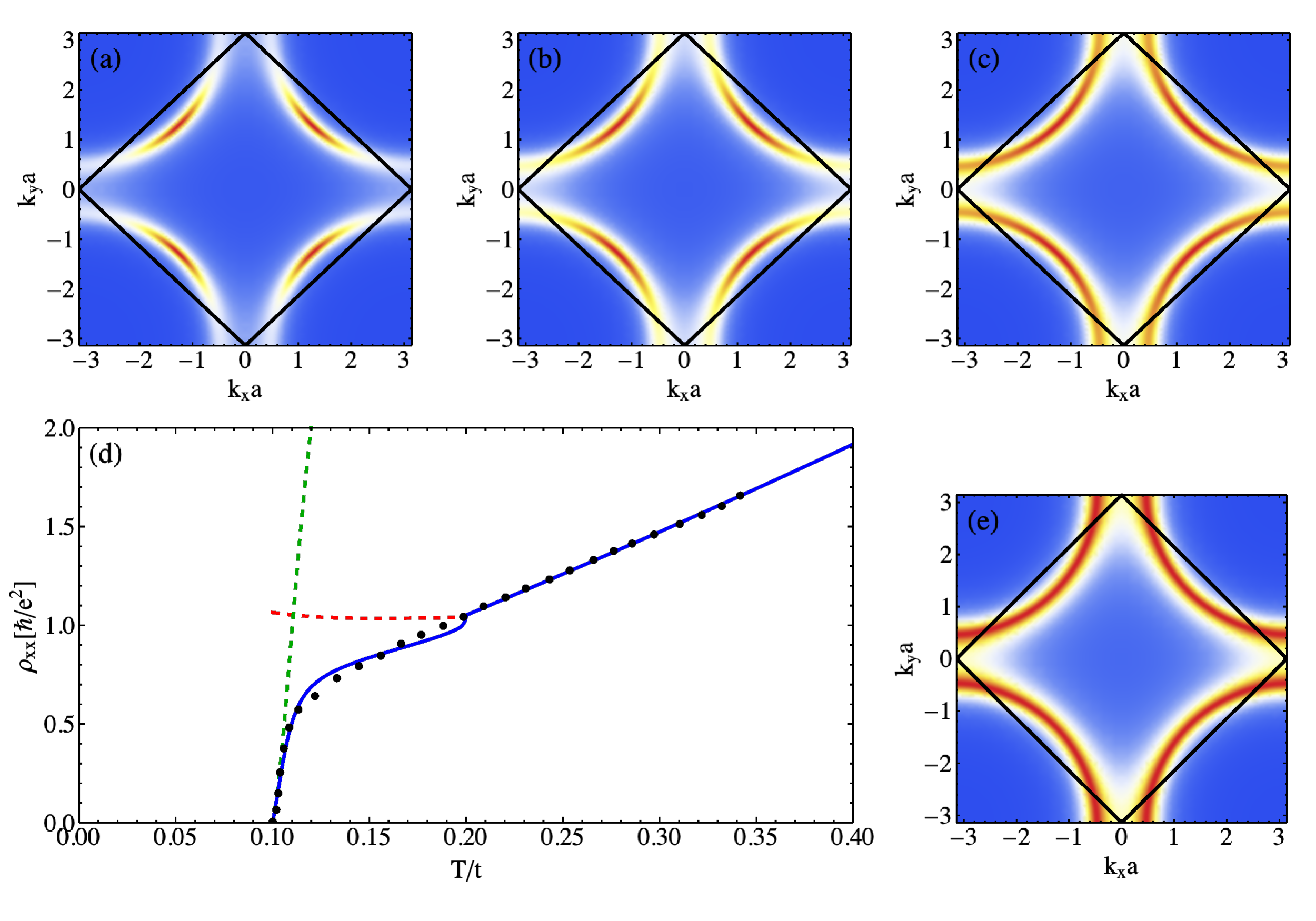}
\caption{Fermi arcs and fit to the resistivity of a slightly underdoped
\BSCCO~sample. Black dots in (d)
are experimental data taken from Ref.~\cite{Watanabe1997}.
The fit in (d) is shown as solid and dashed lines
based on a band structure with the hopping integral $t=75\mathrm{meV}$.
Blue solid line: calculated total $\rho_{xx}$. Red dashed line: fermonic $\rho_{xx}^{\text{f}}$. 
Dark green dashed line: bosonic $\rho_{xx}^{\text{b}}$.
The upturn in $\rho_{xx}^{\text{f}}$ with decreasing $T$, relative to the linear-$T$ background, represents
the fermionic pseudogap signature which is reflected in the fermionic spectral function at the fermionic chemical potential $\mu_F(T)$, $\AF(z=0,\vect{k})$. 
(a-c,e) Representative  $\AF(z=0,\vect{k})$ for temperatures (a) $T/t=0.11$, (b) $T/t=0.15$, (c) $T/t=0.18$, and (e) $T/t=0.23$.
Here, as in all figures, $T_c/t = 0.1$ and $T^*/t = 0.2$.
The evolution shown in these figures shows that the Fermi-arc length, chosen to be reasonably
consistent with photoemission, decreases with decreasing $T$. Note that the small kink at $T^*$ in
the theory should be viewed as an artifact; it arises from the assumed abrupt divergence in the bosonic chemical potential and the onset of the pseudogap.}
\label{fig:Fig1}
\end{figure*}

We divide the following discussion into the temperature ranges above and below $T^*$.
Above $T^*$, to phenomenologically accommodate the linearity of $\rho_{xx}$ in $T$
we use $\Sigma_0(T) =  \Gamma_0 + b T$ for the normal state (fermionic) inverse lifetime in Eq.~\eqref{eq:FermiArcG},
where $\Gamma_0$ and $b$ are two $T$-independent constants. $\Gamma_0$  characterizes the extrapolated residual resistivity. 
$\Gamma_0$ and $b$ are chosen so that $\rho_{xx}$ at $T>T^*$ fits the experimental data of
a nearly optimally doped \BSCCO~ sample in Ref.~\cite{Watanabe1997}, as shown in Fig.~\ref{fig:Fig1}. 
Note that above $T^*$ there is no bosonic contribution to $\rho_{xx}$ so that $\rho_{xx}=\rho_{xx}^{\text{f}}$ with
$\rho_{xx}^{\text{f}}=1/\sigma_{xx}^{\text{f}}$,
where $\sigma_{xx}^{\text{f}}$ is calculated from Eq.~\eqref{eq:Lresp2} using the fermionic spectral function given in Eqs.~\eqref{eq:FermiSpectral} and \eqref{eq:FermiArcG}.
Also, in Eq.~\eqref{eq:FermiArcG}, $\Delta_{\pg}\equiv 0$ for $T\ge T^*$. 
The fitting in Fig.~\ref{fig:Fig1} gives $\Gamma_0/t=0.102$ and $b=1.4$.

Below $T^*$ both the bosonic conductivity $\sigma_{xx}^{\text{b}}$ and  $\Delta_{\text{pg}}$ become nonzero. 
We calculate
\footnote{
For numerical calculations, we choose the integral limit in Eq.~\eqref{eq:Lresp} to be
$\int_{-\pi}^{\pi} \int_{-\pi}^{\pi}  \frac{d^2 p}{(2\pi)^2} \int_{- 5 k_B T}^{5 k_B  T} \frac{dz}{2\pi}$.}
$\sigma_{xx}^{\text{b}}$ from Eqs.~\eqref{eq:Lresp}, \eqref{eq:GG0Prop}, and \eqref{eq:BoseSpectral},
which depend on $\{\kappa,M_{\pair}, \mu_{\pair}\}$ in Eq.~\eqref{eq:Lresp}.
With $\mu_{\pair}$ from Eq.~\eqref{eq:intrplmub} and
$\{\kappa,M_{\pair}\}$ determined from microscopic theory~\cite{QijinThesis,kappanote}
 the only free and adjustable parameter is the inverse lifetime of the pairs, $\Gamma$, in Eq.~\eqref{eq:Lresp}.
We choose the value of $\Gamma$ such that
the total combination of fermionic and bosonic contributions to
$\rho_{xx}$ fits the experimental $\rho_{xx}$ at $T_c \le T \le T^*$ reasonably well
(see Fig.~\ref{fig:Fig1}). 
This leads to $\Gamma=3$.

In Fig.~\ref{fig:Fig1}, the fermionic contribution to the resistivity, $\rho_{xx}^{\text{f}}$ (red dashed line),
deviates from its linear in $T$ background below $T^*$ and exhibits the expected upturn as $T$ decreases.
Countering this upturn we see that the presence of bosons provides another conducting channel, which tends to increase the total $\sigma_{xx}=\sigma_{xx}^{\text{f}}+\sigma_{xx}^{\text{b}}$
and leads to a downward deviation of $\rho_{xx}$ from its high-temperature extrapolation~
\footnote{Notice that in Fig.~\ref{fig:Fig1}, right below $T^*$, the total $\rho_{xx}$ shows a sudden drop which is an artifact of our theory due to the logarithmic divergence
of $\mu_{\pair}$ that we used (see Eq.~\eqref{eq:intrplmub}).}.
Very near $T_c$ the bosonic $\rho_{xx}^{\b}=1/\sigma_{xx}^{\b}$ (green dashed line) is linear in $T-T_c$, as one can see from Eqs.~\eqref{eq:Sigmaxx} and \eqref{eq:intrplmub}~
\footnote{
We note that had we used a different scaling for $\mu_{\pair}$, i.~e., $\mu_{\pair} \propto (T-T_c)^{\alpha}$ with $\alpha$ an integer $> 1$, we would not produce the steep rise of $\rho_{xx}$
at $T\approx T_c$, as seen experimentally. This partially justifies our choice of $\mu_{\pair}$ in Eq.~\eqref{eq:intrplmub}}.

\subsection{Behavior in the conducting channels}

\begin{figure}[h]
\centering
\includegraphics[width=.48\textwidth]
{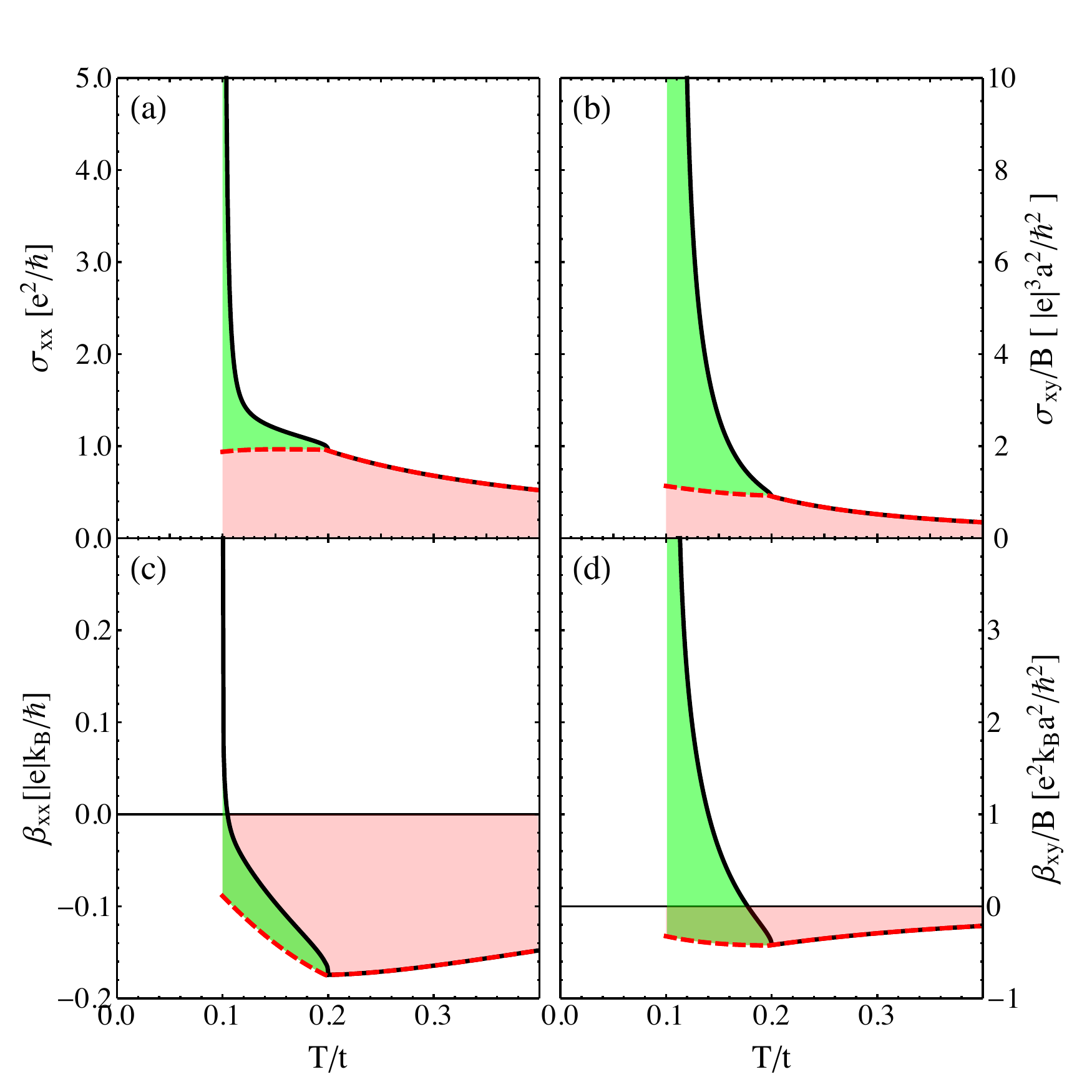}
\caption{Consequences of resistivity fits,
based on Fig.~\ref{fig:Fig1},
as seen in other transport properties.
In each plot, the red dashed line (and the shaded area in red) represent fermionic contributions alone;
while the black solid line represents the total from both fermions and bosons.
The regime shaded in green reflects contributions from fluctuating Cooper pairs.
Here $a$ is the lattice constant, appearing in the units of $\sigma_{xy}/B$ and $\beta_{xy}/B$.
The fermionic contribution is rather small in all properties as a consequence of
the bad-metal character. }
\label{fig:Fig2}
\end{figure}

The fundamental theoretical properties computed in this paper derive from Eq.~\eqref{eq:Lresp} and Eq.~\eqref{eq:Lresp2} 
and pertain to the conducting channel. Once we have established the resistivity fit, there is no parameter flexibility
so that these properties are predetermined. 
Using the $\Sigma_0$ and $\Gamma$ determined from the $\rho_{xx}$ fit in Fig.~\ref{fig:Fig1} we compute all of the other transport coefficients. 
The resulting behavior of $\sigma_{xx}$,  $\sigma_{xy}/B$, $\beta_{xx}$, and $\beta_{xy}/B$
are shown in Fig.~\ref{fig:Fig2}. 

There are some general trends in Fig.~\ref{fig:Fig2} that are rather universal. First, the magnitude of all fermionic 
conduction-like quantities (red dashed lines)
at $T<T^*$ becomes smaller than their high-temperature extrapolations, due to the opening of a pseudogap. 
Second, the bosonic contributions to $\sigma_{xx}, \sigma_{xy}/B, \beta_{xx}$ and $\beta_{xy}/B$ diverge as $T\rightarrow T_c$
from above. 
In particular, the divergence of $\sigma_{xx}^{\b}$ leads to a
noticeable feature in the $\rho_{xx}$ plot of Fig.~\ref{fig:Fig1}: the bosonic contribution (to conductivity) dominates
that of the fermions, albeit in a limited temperature range; for a summary of the divergences of the bosonic transport contributions, see Sec.~\ref{sec:divergence}.

In Fig.~\ref{fig:Fig3} we address other important experimentally accessible transport quantities. 
We have already discussed $\rho_{xx}$, shown in Fig.~\ref{fig:Fig3}(a).
In Figs.~\ref{fig:Fig3}(b)-\ref{fig:Fig3}(d), we plot the Hall coefficient $R_{\H}$, the Seebeck coefficient $S_{xx}$, and the Nernst coefficient $\nu$, based on the same set of parameters.
We have approximated $R_{\H}$, $S_{xx}$, and $\nu$, using Eqs.~\eqref{eq:RH1}, \eqref{eq:Thermo1}, and \eqref{eq:Nernst1}, respectively, as follows 
\begin{align}
\label{eq:6.3} R_{\text{H}}& \approx \frac{1}{B} \frac{\sigma_{xy}}{\sigma_{xx}^2}, \\
S_{xx}&\approx \frac{ \beta_{xx} }{\sigma_{xx}}, \\
\label{eq:Nernst2} \nu&\approx\frac{1}{B}\frac{\beta_{xy}}{\sigma_{xx}}. 
\end{align}
In order to neglect the $\sigma_{xy}^2$ in the denominator of $R_{\text{H}}$ and $\nu$, 
we have restricted our focus to
the weak magnetic field limit and also to $T$ not too close to $T_c$ such that $|\mu_{\pair}| \gtrsim \omega_c^{\text{b}}= 2 e B / M_{\pair}$. 
When this latter condition is satisfied, $|\sigma_{xy}| \ll \sigma_{xx}$.

Some general features of Fig.~\ref{fig:Fig3} are: (i) with the exception of $S_{xx}$ and $\kappa_{xx}$ all other quantities (excluding $\rho_{xx}$) show
divergences or near-divergences in the vicinity of $T_c$ due to their bosonic contributions. 
Both $R_{\text{H}}$ and $\nu$ contain cancelling divergences coming from the numerators and denominators of
Eq.~\eqref{eq:6.3} and Eq.~\eqref{eq:Nernst2}, but nevertheless they lead to strong peaks in transport.
(ii) The bosonic contributions are substantial over a wide temperature range above $T_c$, which can be also seen from Fig.~\ref{fig:Fig2}. 
This derives from the fact that the fermionic conductivities (including $\sigma_{xx},\sigma_{xy},\beta_{xx}, \beta_{xy},\kappa_{xx}$
and $\kappa_{xy}$) are relatively small in cuprates due to their bad-metal character which
in turn derives from the large value for $\Sigma_0$ in Eq.~\eqref{eq:FermiArcG}. 
Importantly, if one considers the case of good metals (see Appendix~\ref{app:GoodBad}) we find the bosonic contributions are confined to the rather narrow critical regime near $T_c$. 
(iii) Related to (ii), the change of the fermionic contributions across $T^*$ due to the onset of the pseudogap is rather weak, 
which will be contrasted with the good-metal case where the change is quite dramatic (see Appendix~\ref{app:GoodBad}). 
Finally, (iv) Regarding the behavior at high temperature above $T^*$, the quantities plotted in Fig.~\ref{fig:Fig3} can be divided into two groups, $\{\rho_{xx}, \nu, \kappa_{xx}/T,\kappa_{xy}/T\}$ and $\{R_{\text{H}}, S_{xx}\}$. 
The former group depends on the inverse fermionic lifetime, $\Sigma_0$.  Consequently, their magnitudes 
are dependent on the fact that the cuprates are bad metals.
In contrast, $R_{\H}$ and $S_{xx}$ do not depend on $\Sigma_0$~
\footnote{For this reason, previous work by our group~\cite{Boyack2019},
which focused on $R_{\H}$ alone, did not make a distinction between the behavior of good and bad metals.}
This is because $R_{\H}$ involves the ratio of $\sigma_{xy}$ and $\sigma_{xx}^2$ ($S_{xx}$ that of $\beta_{xx}$ and $\sigma_{xx}$),  whose dependence
on this lifetime largely cancels each other.

\subsection{Comparison to cuprate experiments }

In this section, we give a summary of comparisons between our calculated transport coefficients 
in Fig.~\ref{fig:Fig3} and experimental measurements on underdoped cuprates (see Appendix~\ref{app:Comparison}).
At the outset, we identify problematic issues concerning the Hall coefficient and the thermopower
which affect all theoretical attempts to understand these cuprate data and make a direct
comparison between theory and experiment difficult.

Indeed, there is a sizable literature dealing with the Hall coefficient in the underdoped regime~\cite{Rice1991,Hwang1994,Lang1994,Samoilov1994,Jin1998,Konstantinovic2000,Matthey2001,Ando2002,Segawa2004}.
Among the most serious problems is that $\sigma_{xy}$ is not as singular near $T_c$ as is predicted by
Gaussian fluctuation theories, where the expected singularity is stronger than in $\sigma_{xx}$ (see Sec.~\ref{sec:divergence}). 
This is presumably associated with the observation that
$R_{\text{H}}\propto \rho_{yx}$ starts to drop with decreasing $T$ at $T$ slightly above $T_c$~\cite{Lang1994,Jin1998} and can even change its sign as $T$ decreases towards $T_c$.
Moreover $R_{\text{H}}$ in the normal state above $T^*$ has a characteristic and systematic $1/T$ dependence~\cite{Clayhold1989,Rice1991}
of unknown origin which serves as a background on top of which paraconductivity and pseudogap effects emerge. 

Similarly, the normal state thermopower in underdoped cuprates~\cite{Munakata1992,Huang1992,Fujii2002,Badoux2016,CyrChoiniere2017}
(at $T\sim T^*$) is positive in the experiments for the samples with the largest pseudogap. 
This is opposite to the band structure predictions with a frequency and $\vect{k}$ independent $\Sigma_0$,
and also opposite to the sign of the Hall coefficient as has been noted previously in Refs.~\cite{Storey2013,Verret2017}.

Given the easily anticipated problems outlined above for $S_{xx}$ and $R_{\text{H}}$, 
comparisons between experiments and our plots are semi-quantitatively reasonable only
for the case of the Nernst coefficient, $\nu$.
Indeed, measurements of $\nu$ on underdoped cuprates~\cite{Xu2000,Wang2001,Wang2006,Chang2011,CyrChoiniere2018} have a long history.
However, there are some non-universalities concerning the Nernst effect, where
there seems to be two classes of behavior.
Both \LSCO\ and \BSCCO\  exhibit a negative contribution to $\nu$ for $T>T^*$, which is to be associated with
the fermions and their band structure. By contrast
\YBCO\  and \HgBCO\  exhibit a positive $\nu$ at $T>T^*$~\cite{CyrChoiniere2018}, inconsistent with
their band structure.

In these latter compounds, $\nu$ experiences two sign changes as $T$ drops below $T^*$.
It changes first from positive to negative, and then back to positive at a lower $T$ near $T_c$.
In Ref.~\cite{CyrChoiniere2018} the first sign change at higher temperature has been taken as evidence against pairing fluctuations playing an important role at $T\approx T^*$.
By contrast, the experimental data of $\nu$ at $T>T_c$ in underdoped La$_{2-x}$Sr$_x$CuO$_4$ and \BSCCO ~\cite{Xu2000,Wang2001,Wang2006,CyrChoiniere2018} is rather similar to that calculated in this paper
and shown in Fig.~\ref{fig:Fig3}(d) (see Appendix~\ref{app:Comparison}).
Before arriving at any conclusions it will be important to
better understand both experimentally and theoretically the non-universal aspects of
the Nernst data observed in the two classes of materials mentioned above.

Finally, we note that the thermal conductivities $\kappa_{xx}$ or $\kappa_{xy}$ do not show any distinctive features as $T$ decreases across $T^*$, 
in agreement with experiments \cite{Krishana1999,Zhang2000}. Also important is the fact that
the bosonic contribution to $\kappa_{xx}$ 
and $\kappa_{xy}$ at $T_c<T<T^*$ in Fig.~\ref{fig:Fig3}(e) and \ref{fig:Fig3}(f) are respectively negligible, or only very weak.
It should be noted that experimentally, at least in $\kappa_{xx}$ and possibly in
$\kappa_{xy}$ (for rather exotic chiral phonons), phononic contributions should play a role
and can mask possible signatures from the charged particles.

We end by discussing to what extent we should view cuprate transport as universal.
The onset of the pseudogap in the resistivity has been shown to be
associated with both an upturn deviation from the linear background as well as a downturn signature.
Here we have looked at the case of a downturn which we interpret as suggesting that the bosonic contribution from 
the pseudogap dominates that coming from the fermions.
We find that a fit to an alternative picture where an upturn is seen from $T^*$ downwards~\cite{CyrChoiniere2018}, for example in \LSCO and related cuprates, 
is possible only if the transition temperatures are rather low; when fitted in this way,
we find that the remaining underlying transport behavior is not substantially changed.
In this case, the fermions will be slightly more prominent in the vicinity of $T^*$.

\begin{widetext}
\begin{center}
\begin{figure}[htp]
\includegraphics[width=0.85\textwidth,trim=0mm 10mm 0mm 10mm]
{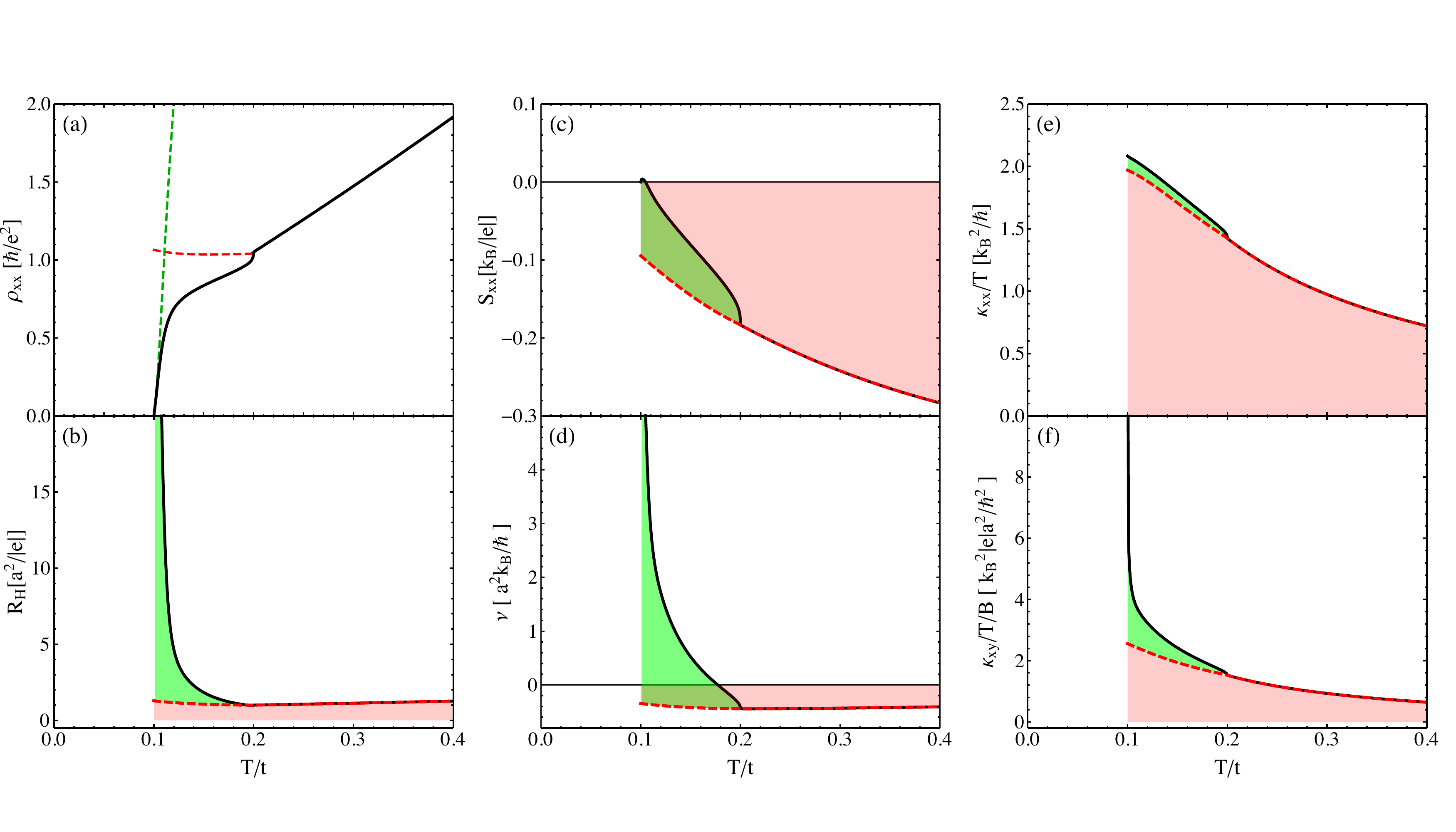}
\caption{Experimentally measurable transport quantities. 
The resistivity plot in (a) is the same as in Fig.~\ref{fig:Fig1}, included for comparison.
In all other plots, the color and line codings are the same as in Fig.~\ref{fig:Fig2}.
In this bad-metal case appropriate to the cuprates, the Cooper-pair fluctuation effects become apparent slightly below
$T^*$.} 
\label{fig:Fig3}
\end{figure}
\end{center}
\end{widetext}

\section{Open-Circuit Contribution}
\label{sec:opencircuit}

An anomalously large and negative value for $\kappa_{xy}$ measured by Ref.~\citep{Grissonnanche2019} has led to substantial theoretical interest in the thermal Hall conductivity. 
The measured thermal conductivity, like the Nernst coefficient in Eq.~\eqref{eq:Nernst1}, is determined under open-circuit conditions. 
As shown in Eq.~\eqref{eq:Jh1}, there are two terms in the expression for $\tens{\widetilde{\kappa}}$ -- an intrinsic contribution arising from $\tens{\kappa}$ and the
open-circuit contribution arising from $\tens{\gamma}\tens{\sigma}^{-1}\tens{\beta}$. 
Here we are interested in the weak magnetic field limit, and so we retain terms in the numerator of Eq.~\eqref{eq:kappayx} only to linear order in the magnetic field and in the denominator we ignore the field dependence.
We also drop the term proportional to $\beta_{xx}^2$, since it is quadratic in the particle-hole symmetry breaking term of the fluctuation propagator, whereas $\kappa_{xy}$ is linear in this term. 
With these assumptions, $\widetilde{\kappa}_{xy}$ is given by
\begin{equation}\label{eq:kappaxy2}
\widetilde{\kappa}_{xy}\approx\kappa_{xy}-2T\frac{\beta_{xx}}{\sigma_{xx}}\beta_{xy}.
\end{equation}

The authors of Ref.~\citep{Kavokin2020} have called attention to the importance of the open-circuit correction, the second term in Eq.~\eqref{eq:kappaxy2},
which has been argued to dominate $\widetilde{\kappa}_{xy}$
\footnote{In calculating the open-circuit term, the authors of Ref.~\citep{Kavokin2020} have identified $\beta_{xy}$
with $dM^{e}_{z}/dT$. The latter leads to a $1/|\mu_{\pair}|^2$ divergence in $\beta_{xy}$.
Here $M^{e}_z$ is the $z$-component of the electric magnetization. 
However, as shown in Eq.~\eqref{eq:Currents}, $dM^{e}_{z}/dT$
must be combined with the microscopic current contribution, $\tens{L}_{e,h}$ in Eq.~\eqref{eq:Currents}, to calculate $\beta_{xy}$. 
Our theory and that in Ref.~\cite{Ussishkin2002} show that in the Gaussian treatment of superconducting
fluctuations the quadratic divergence is exactly cancelled between the two contributions,
leading to a weaker $1/|\mu_{\pair}|$ divergence in $\beta_{xy}$.}.
In this section we present estimates, from the perspective of both our
numerical calculations as well as experimental measurements, of the
open-circuit correction and deduce that it is too
small by about an order of magnitude to
account for the observed sign change in the  experimental results of
Ref.~\cite{Grissonnanche2019}.

\begin{figure}[htp]
\begin{center}
\includegraphics[width=\linewidth,trim=10mm 5mm 0mm 10mm]{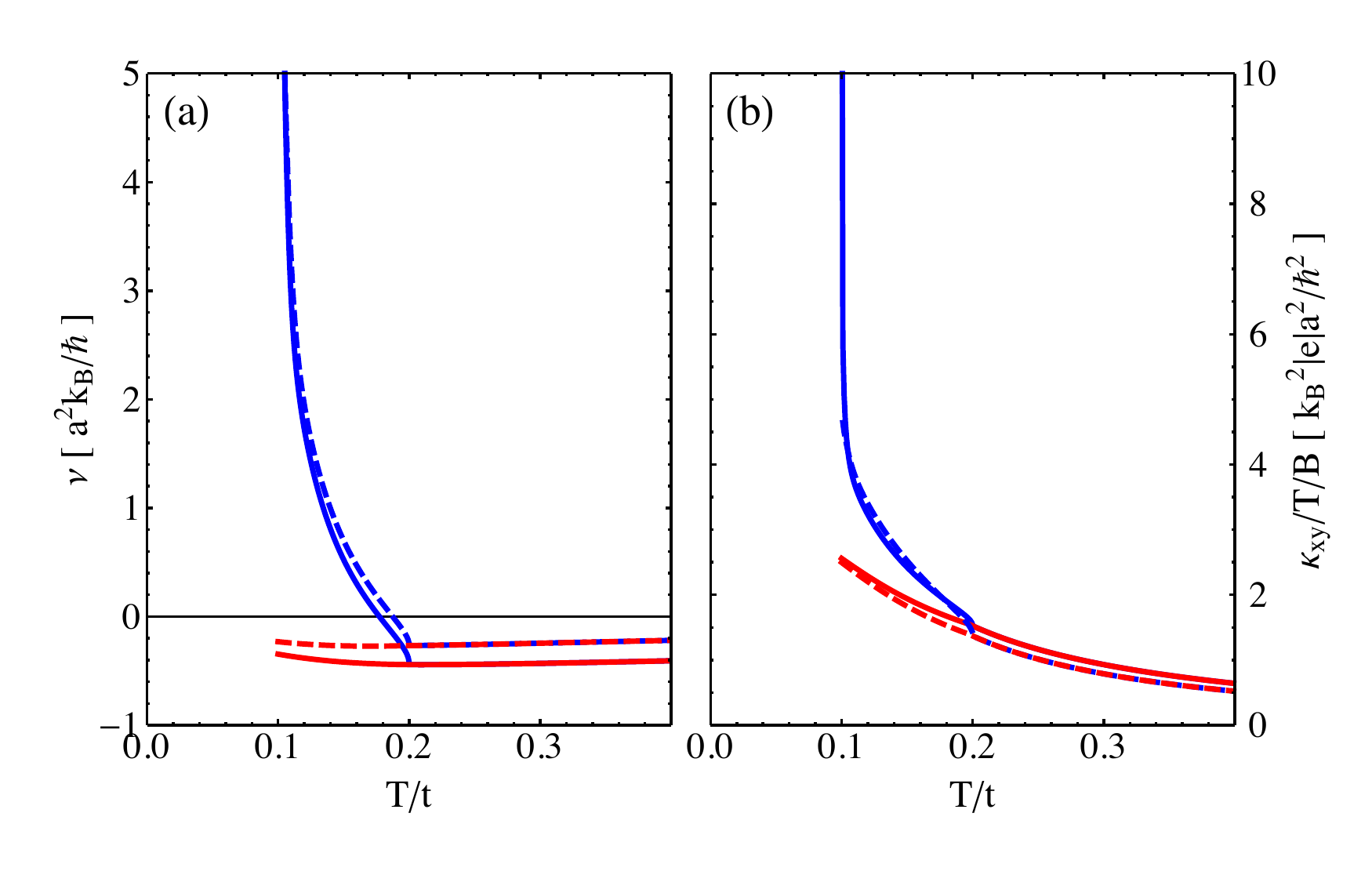}
\caption{Open-circuit correction terms as a comparison between $\{\nu,\kappa_{xy}/T/B\}$ calculated without (solid lines) and with (dashed line) these contributions for
the system of Fig.~\ref{fig:Fig1}.
Red: fermion contributions alone; blue: total contribution from both fermions and bosons.
In both transport properties, the open-circuit corrections are negligibly small. }
\label{fig:Fig4}
\end{center}
\end{figure}

Our numerical results are shown in Fig. \ref{fig:Fig4}. 
The difference associated with the open-circuit terms is reflected in the separation between the solid and dashed lines in this figure. 
Here the blue lines correspond to the total contributions and the red lines to those from the fermionic components. 
We find that the open-circuit terms are negligible except in the very narrow temperature regime when $T$ is very close to $T_c$.
Note that in Fig. \ref{fig:Fig4} the open-circuit correction to $\nu$ is positive.
This is because $\beta_{xx}$ is primarily negative (as can be observed in Eq.~\eqref{eq:Nernst1} and Fig.~\ref{fig:Fig2}).

We next provide numerical estimates based on experimental data from the cuprates.
We define $\Delta\kappa=\tens{\widetilde{\kappa}}-\tens{\kappa}$. From Eq.~\eqref{eq:kappaxy2}
\begin{align} \label{eq:kappaxy3}
\frac{\Delta\kappa_{xy}}{T}
& \approx -2S_{xx}\beta_{xy},
\end{align}
where we have used $S_{xx}\approx\beta_{xx} \rho_{xx}$. 
The value of $S_{xx}$ and $\beta_{xy}$ can be extracted from published experimental data
for \LSCO at hole doping $x\approx 0.07$, which is close to the doping level $x=0.06$ of one sample studied
in Ref.~\cite{Grissonnanche2019}. 
From Ref.~\cite{Badoux2016} we obtain $S/T= 1.0 \,  \mathrm{\mu V/K^2}$. 
From Fig. 4 of Ref.~\cite{Wang2001} we infer $\beta_{xy} \rho_{xx} / B \approx 200  \, \mathrm{ n V/(K\cdot T)} $.
Using $\rho_{xx}=0.3 \, \mathrm{ m\Omega\cdot cm }$ for $x=0.08$ from Ref.~\cite{Ono2007}, we obtain $\beta_{xy}/B= 66 \, \mathrm{ m A / (m\cdot K \cdot T) }$. 
We use the $x=0.08$ $\rho_{xx}$ data for our estimate 
since we do not expect a significant change of  $\rho_{xx}$ from $x=0.08$ to $x=0.07$.
Inserting these values for $S_{xx}$ and $\beta_{xy}$ into Eq.~\eqref{eq:kappaxy3} for $B=15$T and $T=30$K leads to
\begin{equation}
\frac{\Delta \kappa_{xy}}{T}=-0.06 \frac{\textrm{mW}}{\textrm{m}\cdot \textrm{K}^{2}}.
\end{equation}
Comparing with the corresponding experimental data~\cite{Grissonnanche2019} for $x=0.06$ at $B=15$ T and $T=30$K, this appears to be too small by more than one order of magnitude to be relevant~
\footnote{Here we consider the case where the applied $B$ field is small compared to $B_{c2}$. 
At larger fields, an estimate for the neglected terms shows that they are even smaller than $2 S_{xx} \beta_{xy}$. 
We note that our open-circuit estimate in this section does not rely on the weak-field limit, although our numerical result in Sec.~\ref{sec:Numerics} does.}.

\section{Conclusions}
\label{sec:Conclusions}

In this paper we have shown how to address the broad class of fermionic and bosonic
transport coefficients in a consolidated fashion in the linear magnetic field regime.
Our theory is based on a reservoir approach developed for non-interacting particles.
As an illustration, we applied these ideas to the normal state of the superconducting cuprates,
under the hypothesis that the pseudogap phase consists simultaneously of both (gapped) fermions and bosonic pairs.
This paper deals with the challenge of looking at a wide array of transport
coefficients in superconductors with such a pseudogap.
The challenge comes from the fact that the two types of charge carriers can have
competing or enhancing contributions.
Understanding which of these dominates and in which experiment is an important goal of our paper.

For the boson channel
there is a general consensus that these Aslamazov-Larkin-like contributions
can be modelled as essentially independent bosons. 
Thus the central formulae of this paper [Eq.~\eqref{eq:Lresp}] are on a rather firm footing.
The situation is more complicated with respect to the fermions, where the independent
particle-assumption is only an approximation.
In the applications to the cuprates involving fermionic contributions, while we cannot argue that we have fully
incorporated the Ward identities or gauge invariance, we do include interaction effects through the pseudogap self energy term. 
This is responsible for the important Fermi arc effects.
As shown in Ref.~\cite{Scherpelz2014}, the neglected vertex corrections (for the longitudinal response) are relatively small.

Most importantly, despite our more approximate treatment of the fermionic
contributions, one can see from Fig.~\ref{fig:Fig2} that the
dominant features to the transport
coefficients are the bosonic contributions, which generally have
a tendency to diverge or become very large.
That they are so apparent over a wide range of
temperatures is an important conclusion of this paper. 
We associate this with the bad-metal character of the cuprates which allows the bosons to be more dominant.
Thus, because of this bad-metal character we do not expect a full many-body approach to qualitatively change our conclusions about transport in the cuprates. 

In the process of looking at the broad class of transport coefficients, 
we have quantitatively studied the experimentally observed behavior of 
the resistivity of a prototypical underdoped cuprate. We consider the 
entire range of temperatures from $T^*$ to $T_c$, assuming that the 
resistivity derives from both fermionic and fluctuating Cooper pair 
(bosonic) contributions. Our goal was to attribute different features in 
the temperature dependence to each of these two sources for charge 
transport and in the process provide information about the origin of the 
pseudogap. In the vicinity of $T_c$ there is no question that the 
behavior is dominated by fluctuating Cooper pairs. Around $T^*$, 
however, there is a notably subtle feature in the resistivity data, usually a slight downturn 
from the linear background. We view the observation that this is so 
slight as very important.
 
Generally we can presume that the $T^*$ signature involves
both bosonic and fermionic contributions. 
Because the latter arise from the opening of a gap in the excitation
spectrum, it is not difficult to anticipate that rather dramatic effects could be evident at $T^*$.
Indeed, calculations in this paper suggest that it is rather challenging (see Appendices) to avoid these abrupt changes
in the resistivity; it is equally problematic to
arrive at~\textit{improved} conductivity (relative to the linear background)
just below $T^*$, when this is associated with 
the onset of a fermionic excitation gap.

\textit{How do we understand the resistivity data, then?} 
In our work we used only one adjustable parameter to fit the resistivity data and found that
one can indeed recover a subtle downturn feature at $T^*$, but only when two key 
and well-known aspects of
the cuprates are included: they are highly resistive or bad metals and the opening of the gap is itself rather subtle
and associated with $d$-wave Fermi arcs.
Also important is the fact that Cooper-pair fluctuations must persist, albeit weakly, up to $T^*$.
It is important, however, to distinguish these from critical fluctuations which we find are 
present only very close to $T_c$.

\textit{What does this indicate about the origin of a pseudogap?} A key 
finding is that this behavior in the resistivity makes it difficult to 
contemplate substantial changes in the fermionic spectral function 
associated with $T^*$. It is not unreasonable to assume that this quite 
possibly rules out new forms of order or Fermi surface reconstructions. 
Rather it suggests that $T^*$ is associated with the onset of some form 
of fluctuating order. Indeed, this is consistent with inferences from 
thermodynamics where there are little or no indications of a true phase 
transition at $T^*$~\cite{Timusk1999}. The presence of fluctuating order associated with 
the pseudogap suggests that bosonic degrees of freedom may be present 
and contribute to transport features at and below $T^*$. If the 
fluctuations are in the particle-particle channel and thus charged, this 
leads to a similar set of complications as was discussed in this paper.

In summary, there is a growing sense that understanding the full complement of thermoelectric transport properties may
shed light on the still-controversial origin of the cuprate pseudogap.
In contrast to the low-field limit we consider, recent emphasis has been on ultra-high magnetic field phenomena where the superconductivity is driven away
but vestiges of the pseudogap in the normal state are presumed to persist, now down to temperature $T=0$.
It remains to be seen whether this pseudogap ground state does or does not
reveal the pristine normal state of the superconducting materials.
Nevertheless, a clear implication is that it is important for theories
to address, as we do here, the broad class of transport properties, and not just a selected few.

\section{acknowledgments}

We are grateful to A. A. Varlamov for beneficial discussions and for sharing Ref.~\cite{Obraztsov1964} with us.
We thank W. Witczak-Krempa and C. Panagopoulos for useful discussions. 
R. B. was supported by D\'epartement de physique, Universit\'e de Montr\'eal. 
Q. C. was supported by NSF of China (Grant No. 11774309).
This work was also supported by the University of Chicago Materials Research Science and Engineering Center, funded by the National Science Foundation under Grant No. DMR-1420709 (K. L. and Z. W.).
It  was completed in part with resources provided by the University of Chicago's Research Computing Center.

\appendix
\numberwithin{equation}{section}
\numberwithin{figure}{section}

\section{Transport in conventional superconducting fluctuation theory}
\label{app:GL}

In this appendix we give a comparison between the conventional GL fluctuation theory in the normal-state of a superconductor~\cite{VarlamovBook}
and the strong-pairing fluctuation theory of Sec.~\ref{sec:framework}. In addition, we provide a brief review of transport literature in the GL fluctuation theory. 

\subsection{Fluctuation propagator}

We first outline how the formula given in Eq.~\eqref{eq:Lresp} can be applied to the case of the GL fluctuation theory~\citep{VarlamovBook}.
This bosonic transport encapsulates the contribution from fluctuating Cooper pairs, and in a diagrammatic framework it corresponds to the Aslamazov-Larkin (AL) fluctuation diagram~\cite{VarlamovBook}.
This approach is traditionally based on Gaussian fluctuations in the normal-state, and as a result it does not directly incorporate a normal-state pseudogap. 

In the GL fluctuation theory~\citep{VarlamovBook}, the inverse fluctuation propagator is defined by 
\begin{equation}
\label{eq:tmat0}
t^{-1}_{0}(p)=\sum_{k}G_{0}(k)G_{0}(p-k)+g^{-1}.
\end{equation}
Here, $p=(i\Omega_{m},\vect{p})$ with $\Omega_{m}$ a bosonic Matsubara frequency. 
The inverse propagator can also be expressed as $t_{0}^{-1}(p)=\Pi(p)+g^{-1}$, where $\Pi(p)$ is the pair susceptibility. 
The small-momentum expansion of the retarded fluctuation propagator is~\citep{VarlamovBook}:
\begin{equation}
\label{eq:GLProp}
t^{-1}_{0,\R}(z,\vect{p}) = iz\gamma_{\GL}-\epsilon_{\vect{p}}.
\end{equation}
Here, $\gamma_{\GL}=\gamma_{1}+i\gamma_{2}$ is the GL parameter and $\epsilon_{\vect{p}}$ is the dispersion relation for fluctuating Cooper pairs. 
In GL fluctuation theory, $\gamma_{2}\sim T_{c}/E_{F}$~\cite{VarlamovBook} where $E_F$ is the Fermi energy, and particle-hole symmetry is only weakly broken.  
As a result, the transport coefficients $\beta_{xx}, \sigma_{xy},$ and $\kappa_{xy}$, which are proportional to $\gamma_{2}$, have a small prefactor. 

The parameters in Eq.~\eqref{eq:GLProp} are given by~\cite{VarlamovBook}
\begin{align}
\gamma_{1}&=\frac{\pi N_F}{8T_{c}}, \\
\label{eq:gamma2}\gamma_{2}&=-\frac{1}{2}N_F\left(\frac{\partial\ln T_{c}}{\partial E}\right)_{E=E_{F}}, \\
\epsilon_{\vect{p}}&=N_F\left(\epsilon+\eta \vect{p}^2\right).
\end{align}
Here, the single-spin density of states at $E_F$ is denoted by $N_F$, $\epsilon=\ln(T/T_c)\approx\left(T-T_{c}\right)/T_{c}$, and $\eta=7\zeta(3)/(16d)\left[\hbar v_{F}/(\pi k_{B}T_{c})\right]^{2}$ for ultraclean systems in $d$ spatial dimensions, where $v_{F}$ is the Fermi velocity. 
The coherence length, $\xi_{0}$, is related to $\eta$ by $\xi^{2}_{0}=\eta$; the temperature-dependent coherence length is $\xi(T)=\xi_{0}/\sqrt{\epsilon}$. 
Note that, $\xi(T)$ is defined in this manner because the fluctuation regime near the critical temperature is of primary concern. 
In particular, the above definitions should be distinguished from the zero-temperature coherence length in BCS theory: 
$\xi_{\text{BCS}}=\hbar v_{F}/(\pi \Delta_{0})$ is the zero-temperature BCS coherence length and $\xi_{0}\approx0.74\xi_{\text{BCS}}$ is the fluctuation coherence length, for an ultaclean three-dimensional system. 
In summary, all of the bosonic transport contributions in GL fluctuation theory can be determined by using Eq.~\eqref{eq:Lresp}, the fluctuation propagator in Eq.~\eqref{eq:GLProp}, and the spectral function
\begin{equation}
\AB(z,\vect{p})=\text{Re}[2it_{0,\R}(z,\vect{p})].
\end{equation}

We end by summarizing the relationship between Eq.~\eqref{eq:GG0Prop} and Eq.~\eqref{eq:GLProp} as follows: 
\begin{align}
\gamma_{1}=Z\Gamma&;\ \gamma_{2}=-Z\kappa,\\
\label{eq:Zmup} N_F \epsilon=Z|\mu_{\pair}|&;\ N_F \eta=Z/(2M_{\pair}).
\end{align}
We relate the coherence lengths via $1/(2k_{B}T_{c}M_{\pair})\rightarrow\eta$, and $|\mu_{\pair}|/(k_{B}T_{c})\rightarrow\epsilon$.
As a consequence, one can deduce the central results for the transport contributions arising from Eq.~\eqref{eq:Lresp}, for both GL and strong-pairing fluctuation theories, by mapping the appropriate terms.

\subsection{Literature summary of transport results}

For the benefit of the reader, here we provide a brief list of the most pertinent literature on the transport coefficients in the GL fluctuation theory.
In addition to the bosonic transport of fluctuation pairs, the normal-state fluctuation theory of a superconductor contains fermionic contributions known as the Maki-Thompson (MT) and Density of States (DOS) terms~\citep{VarlamovBook}.
The formation of fluctuating Cooper pairs causes a decrease in the density of fermions, which gives rise to the DOS term, and it also causes scattering of electrons, which is representative of the MT term.

The AL contribution to the Nernst effect, which dominates as $T\rightarrow T_{c}$, was computed in Refs.~\citep{Ussishkin2003,Serbyn2009}, 
and similarly the contribution to the thermopower was considered by Maki~\cite{Maki1974}.
The AL, MT, and DOS contributions to the longitudinal thermal conductivity were computed in Ref.~\citep{Niven2002}, while the AL contribution was originally considered in Ref.~\citep{Abrahams1970}.
The literature on the electrical conductivity is even more extensive; the MT and DOS diagrams were originally considered in Ref.~\citep{Maki1968} and simultaneously the AL, MT, and DOS diagrams were independently studied in Ref.~\citep{Aslamazov1968}.
For completeness we note that the diamagnetic susceptibility was originally studied in Ref.~\citep{Aslamazov1975} and only recently the shear viscosity has been investigated in Ref.~\citep{Liao2019}.
A complete set of references can be found in Refs.~\citep{VarlamovBook,Varlamov2018}.

Let us now turn to the fluctuation results for the intrinsic thermal conductivity, in the low-field limit,
where there has been some initial controversy surrounding the longitudinal
contribution and where the transverse contribution is more subtle. 
The first fluctuation calculation of longitudinal thermal conductivity was performed by Abrahams et al.~\cite{Abrahams1970}.
These authors noted that the AL diagram ``corresponds to the contribution of the superfluid flow to the current''. 
Since superfluid flow produces no entropy~\citep{Luttinger1964b} it does not transport any heat, and consequently Abrahams et al. concluded that, as the critical temperature is approached, 
the AL diagram is expected to have zero longitudinal thermal conductivity. 
As a result, the main focus of these authors was the thermal response of the DOS and MT diagrams.

Later fluctuation literature~\citep{Varlamov1990,Varlamov1991,Varlamov1992} erroneously concluded, due to a mistreatment of the heat vertex, that the longitudinal fluctuation thermal conductivity is singular.
In confirmation of the result in Ref.~\cite{Abrahams1970}, a hydrodynamic analysis~\citep{Vishveshwara2001}
argued that thermal fluctuations have a nonsingular $\kappa_{xx}$.
A complete and correct microscopic calculation of $\kappa_{xx}$ by Niven and Smith~\citep{Niven2002} ultimately
showed that the singular contributions in the MT and DOS diagrams cancel and the AL diagram itself is non-singular, in arbitrary dimensions and for arbitrary strengths of impurity scattering.

In subsequent work, Ussishkin et al.~\citep{Ussishkin2002} correctly summarized the nature of the
divergences in $\kappa_{xy}$ in GL fluctuation theory,
and, in agreement with Ref.~\citep{Niven2002}, they report that $\kappa_{xx}$ is nonsingular in two and three dimensions.
We emphasize that these results and the final conclusions related to
other transport coefficients are consistent with those of the present paper, as discussed in Sec.~\ref{sec:divergence}.
The derivation of transverse thermoelectric and transverse thermal responses requires the inclusion of magnetization currents.

While there have been numerous publications devoted to particular transport coefficients in the GL theory, there are very few papers with a unified discussion of all the transport coefficients.
In Ref.~\citep{Ussishkin2002}, there is a table of the $d=2$ and $d=3$ results for $\kappa$ and $\beta$ (denoted by $\alpha$ in this reference).
The review in Ref.~\cite{Varlamov2018} discusses electrical and thermoelectric conductivities, but it does not discuss thermal conductivity fluctuation results. 
In this paper it is emphasized that the underlying structure of the bosonic contributions to all transport coefficients in
both conventional and strong-pairing fluctuation theory can be put into the
unified form contained in a single transport equation, Eq.~\eqref{eq:Lresp}, for ultraclean systems.

\section{Detailed comparison between bad and good metals}
\label{app:GoodBad}

In this appendix we discuss the contrast between the cases of good and bad metals.
This is done by varying the size of the underlying linear contribution to
the resistivity, through $\Sigma_0$, for the purpose of showing what happens when
the relative weight of the fermionic and bosonic contributions is changed.
What is more notable in the good-metal case is that (even with Fermi arcs still
present) there are now abrupt features in the fermionic
contributions to transport setting in at $T^*$, which are in contrast to
the relatively subtle features seen in experiment. Additionally, the
bosonic contribution is now restricted to the more conventional
critical regime, around $T_c$.
These calculations are pedagogical, and meant to assist in
understanding the more physical example of a bad metal in the main text.

\begin{figure}[h]
\centering
\includegraphics[width=.48\textwidth]
{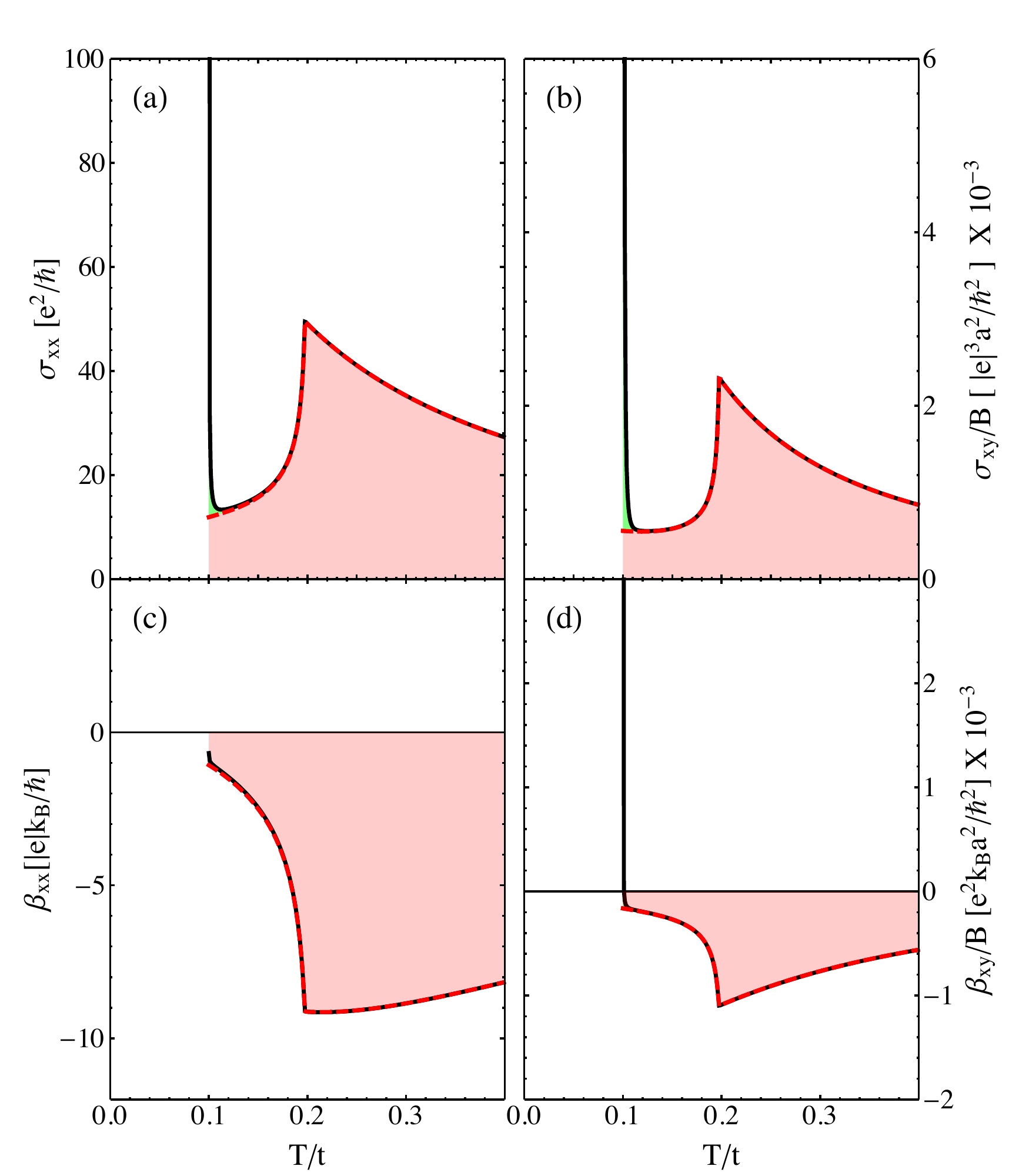}
\caption{Conduction properties in the good-metal case.
In each plot, the red dashed line and the shaded area in red represent fermionic contributions alone;
while the black solid line represents the total contribution from both fermions and bosons.
The regime shaded in green stands for contributions from bosons.
The fermions are quite prominent in the good-metal case.}
\label{fig:Fig5}
\end{figure}

\begin{figure*}
\includegraphics[width=0.85\textwidth,trim=0mm 10mm 0mm 20mm]
{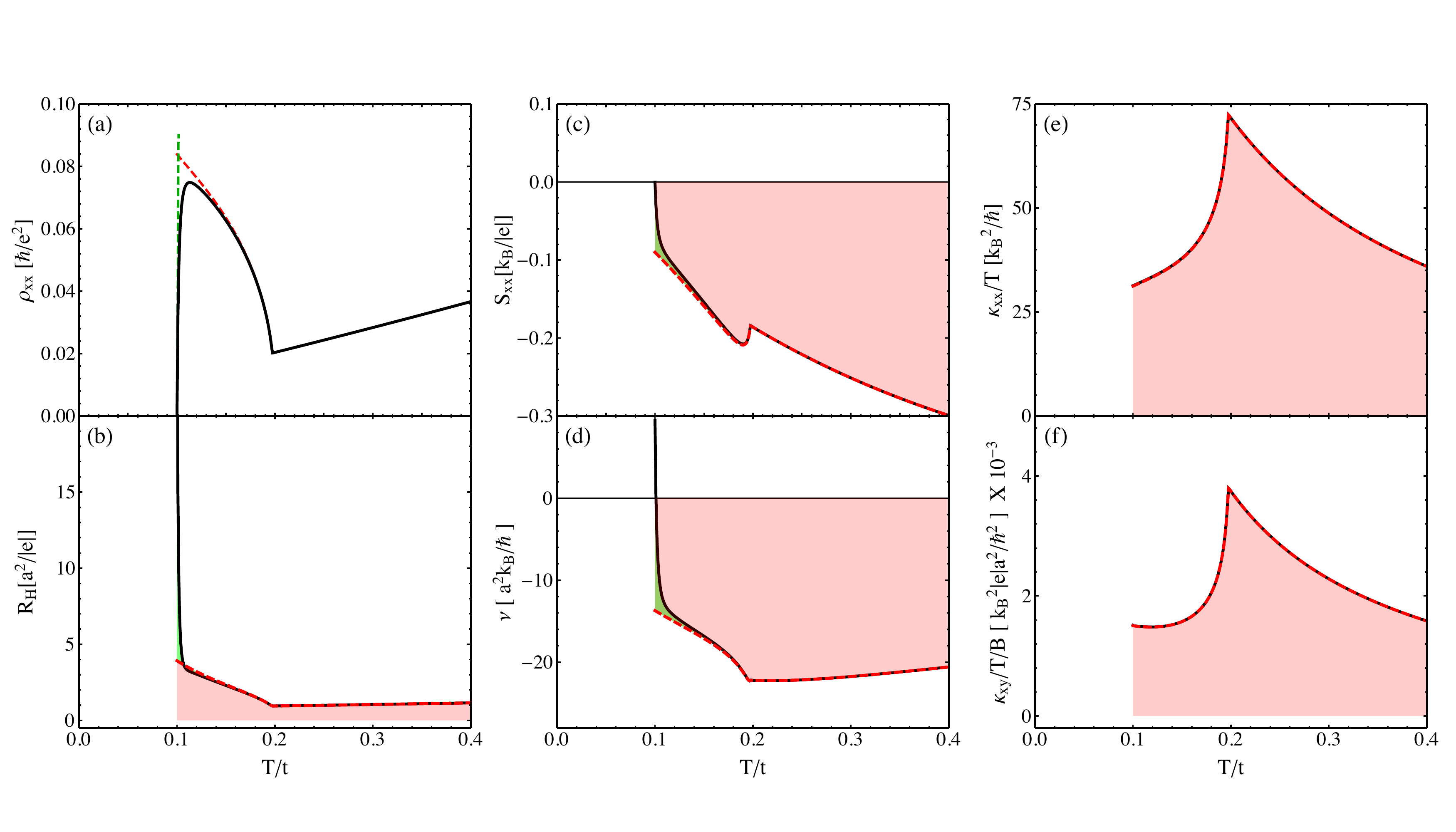}
\caption{Consolidated results for transport: Good-metal case.
In all plots, the color and line codings are the same as in Fig.~\ref{fig:Fig2}.
The upturn in the resistivity for the good metal near $T_c$ is driven by the fact that the
fermions experience a pseudogap in their excitation spectrum and that the effective fermion scattering rate is significantly enhanced below $T^*$ due to the Fermi arc effect (see text). }
\label{fig:Fig6}
\end{figure*}

Indeed, the results for the bad-metal case are already presented in Figs.~\ref{fig:Fig2} and \ref{fig:Fig3};
while those for the good metal are shown in Figs.~\ref{fig:Fig5} and \ref{fig:Fig6}. 
In our calculation, the good metal differs from the bad one by
a 50-fold reduction of the normal-state scattering rate, $\Sigma_0$ in Eq.~\eqref{eq:FermiArcG}, while all other
parameters as well as their temperature dependences remain the same~
\footnote{We note that for good metals, $\rho_{xx}$ typically saturates to a value smaller than the Mott-Ioffe-Regel limit at high enough temperature~\cite{Gunnarsson2003},
which is violated by our good-metal model since $\Sigma_0=\Gamma_0 + b T$ does not saturate,
leading to an unsaturated $\rho_{xx}$. However, this is not our concern because we focus on the relatively low $T_c<T<T^*$.}.
This means that the absolute value of the bosonic contribution is the same for both cases.  
 
Comparing Fig.~\ref{fig:Fig5} to Fig.~\ref{fig:Fig2}, we observe that one of the most important distinctions is that, in the 
physically more relevant bad-metal case, the bosonic contribution
can be substantial even at relatively high temperatures near $T^*$, whereas its relevance is highly restricted to a narrow temperature range near $T_c$ in the good-metal case.
This is because from the bad to good metal, the bosonic contribution does not change while the fermionic 
counterpart increases by about $50$ (for quantities such as $\sigma_{xx}^{\text{f}}$)
or $50^2$ (for quantities such as $\sigma_{xy}^{\text{f}}/B$) due to the decrease of $\Sigma_0$ in Eq.~\eqref{eq:FermiArcG}.
We note that, in Fig.~\ref{fig:Fig5}(c), $\beta_{xx}$ is expected to diverge logarithmically at
$T=T_c$ (see Sec.~\ref{sec:divergence}), which is, however, cut off by finite-size effects in our numerical calculation.

Another distinction between Fig.~\ref{fig:Fig5} and Fig.~\ref{fig:Fig2} is that, in the good-metal case, the magnitude of all of the fermionic conductivities, 
$\{\sigma_{xx}^{\text{f}}, \sigma_{xy}^{\text{f}},\beta_{xx}^{\text{f}},\beta_{xy}^{\text{f}} \}$ (and also $\{\kappa_{xx}^{\text{f}}/T, \kappa_{xy}^{\text{f}}/T/B \}$
in Fig.~\ref{fig:Fig6}),
drops rapidly as $T$ lowers below $T^*$, resulting in a cusp-like feature at $T^*$. This is in sharp contrast to the bad-metal case
where the change in $T$ dependence of fermionic conductivities across $T^*$ is rather weak. 
The sharp drop in the good-metal case originates from an increase in the effective fermion scattering rate, which changes from $\Sigma_0 $
at $T>T^*$ to $\Sigma_{0,\text{eff.}} = \Sigma_0 + ( \Delta_{\text{pg}}(T) \varphi_{\vect{k}} )^2/ \gamma$ (see Eq.~\eqref{eq:FermiArcG}; here, we consider $\vect{k}=\vect{k}_F$ and $\omega=0$). 
Even just slightly below $T^*$, $\Sigma_{0,\text{eff.}}  \gg  \Sigma_0$ because $\Delta_{\text{pg}} \sim \gamma \gg \Sigma_0$.
This rapid increase in scattering rate leads to the rapid drop of all conductivities below $T^*$. 
The effect of scattering rate change across $T^*$ is much weaker for bad metals because there $\Sigma_0 \sim \Delta_{\text{pg}}\sim \gamma$ are all
comparable. 

We turn now to Fig.~\ref{fig:Fig6}.
One notable feature is that the upturn of the fermionic $\rho_{xx}^{\text{f}}$ at $T<T^*$
in Fig.~\ref{fig:Fig6}(a) is more pronounced in the good-metal case, as compared to Fig.~\ref{fig:Fig3}(a).
The sharp upturn corresponds to the rapid drop of $\sigma_{xx}^{\text{f}}$ in Fig.~\ref{fig:Fig5} right below $T^*$. 

The behaviors of $R_{\text{H}}, S_{xx}$, and $\nu$ in Fig.~\ref{fig:Fig6} are quite similar to those in Fig.~\ref{fig:Fig3} except that, for good metals,
(1) the fermionic contribution to $R_{\text{H}}$ has a more pronounced upturn at $T<T^*$, because $R_{\text{H}}^{\text{f}} \propto 1/ n_{\text{f}}$
and the effect of loss of fermionic carrier density $n_{\text{f}}$ is stronger,
and (2) the magnitude of $\nu$ is much larger because $\nu^{\text{f}} \propto 1/\Sigma_0$. 
The similarity comes from the fact $S_{xx}^{\text{f}}$ and $R_{\text{H}}^{\text{f}}$ are rather
insensitive to $\Sigma_0$. 

Another similarity between Fig.~\ref{fig:Fig6} and Fig.~\ref{fig:Fig3} 
is that in both cases the bosonic $\kappa_{xx}^{\text{b}}$ is negligible, 
which follows because $\kappa_{xx}^{\text{b}}$ does not diverge as $T\rightarrow T_c$ in 2d (see Sec.~\ref{sec:divergence}).
This is in contrast to the bosonic contribution
to $\kappa_{xy}^{\text{b}}$ which shows a weak logarithmic
divergence as $T\rightarrow T_c$. This divergence is not
visible in Fig.~\ref{fig:Fig6}(f) due to 
finite-size effects in our numerical calculations.
Note also that in Fig.~\ref{fig:Fig6}, the fermionic contributions
to $\kappa_{xx}/T$ and $\kappa_{xy}/T$ are so large that the bosonic contributions become almost invisible.

Overall, the comparisons between Fig.~\ref{fig:Fig3} and Fig.~\ref{fig:Fig6} again underline
the fact that the bosonic contributions are much more prominent over a large temperature range above $T_c$ in the 
more physical case of a bad metal.

\section{Detailed comparison between cuprate data and our theory}
\label{app:Comparison}

In this section we focus on qualitative temperature dependence of the transport quantities
while deferring a brief discussion on their magnitudes to Appendix~\ref{app:unit}. 

\subsection{Hall Coefficient} 

We start with the Hall coefficient. There is a large body of Hall measurements on underdoped cuprates~\cite{Rice1991,Hwang1994,Lang1994,Samoilov1994,Jin1998,Konstantinovic2000,Matthey2001,Ando2002,Segawa2004,Doiron-Leyraud2007,Badoux2016a}, focusing on different hole doping, temperature, and magnetic-field regimes. The Hall coefficient measured on moderately hole-doped \YBCO\ at low temperature and high magnetic field exhibits pronounced quantum oscillations~\cite{Doiron-Leyraud2007} with a small oscillation frequency $F\sim 530 $ Tesla, which corresponds to a Fermi surface area only about 2$\%$ of the Brillouin zone and suggests that the bare large hole-like Fermi surface gets reconstructed. 

In this paper, we focus on the weak magnetic field and high temperature ($T>T_c$) limit, where we assume no such reconstructions.
In this limit, the Hall coefficient, $R_{\text{H}}$, measured on underdoped cuprates shows a well known $1/T$ dependent background~\cite{Rice1991,Clayhold1989}, whose origin remains undetermined.
One explanation~\cite{AndersonBook} presumes two distinct normal-state lifetimes, one for $\sigma_{xx}^{\text{f}}$ and another for $\sigma_{xy}^{\text{f}}$.
In our calculations we do not consider such a distinction; instead, we use the same $\Sigma_0$ in Eq.~\eqref{eq:FermiArcG} for both $\sigma_{xx}^{\text{f}}$
and $\sigma_{xy}^{\text{f}}$. Consequently, our calculated $R_{\text{H}}$ in Fig.~\ref{fig:Fig3}(b) is essentially temperature independent at $T>T^*$, as expected in a single-lifetime scenario.

The other key feature of the low field Hall data on underdoped cuprates is that, below some characteristic $T$ slightly above $T_c$, $R_{\text{H}}$ decreases with decreasing $T$ ~\cite{Lang1994,Jin1998} and can even change its sign as $T$ drops below $T_c$,
in sharp contrast to Fig.~\ref{fig:Fig3}(b) where $R_{\text{H}}$ continues to grow as $T\rightarrow T_c$ from above.
Very near $T_c$ our calculated $R_{\text{H}}$ eventually saturates because $R_{\text{H}}$ is dominated by the bosonic
contribution, and it is $\propto \rho^{\text{b}}_{yx} \propto \sigma^{\text{b}}_{xy} / (\sigma_{xx}^{\text{b}})^2$, where the divergences of $\sigma^{\text{b}}_{xy}$
and $(\sigma_{xx}^{\text{b}})^2$ cancel out each other (see Sec.~\ref{sec:divergence}). 
This saturation is not shown in Fig.~\ref{fig:Fig3}(b) because our theory is valid only for  $|\mu_{\pair}| \propto |T-T_c| \gtrsim 2 e B/M_{\pair}$ ($eB>0$). 
Experimentally, the downturn of $R_{\text{H}}$ seems to come from $\sigma_{xy}^{\text{b}}$ displaying a weaker singularity
than $\sigma_{xx}^{\b}$ as $T\rightarrow T_c$~\cite{Rice1991,Jin1998}.

An alternative way to reconcile theory with experiments is to assume that the divergent part of the bosonic $\sigma_{xy}^{\text{b}}$ 
carries a sign opposite to that of $\sigma_{xy}^{\text{f}}$~\cite{Rice1991}.
However, we emphasize that this sign, dictated by the particle-hole asymmetry factor $\kappa$ in our theory (see Sec.~\ref{sec:divergence}),
is not arbitrary but correlated with the underlying fermionic band structure that gives rise to Cooper pairs.
For the band structure we use, $\kappa$ is found to be negative in Ref.~\cite{Boyack2019},
leading to the positive $\sigma_{xy}^{\text{b}}$ in Fig.~\ref{fig:Fig2}(b). 

We could hypothesize that some additional physics due to vortices (not included in our theory), such as discussed in Ref.~\cite{Auerbach2020}, can lead to a negative
$\sigma_{xy}^{\text{b}}$ and account for the downturn of $R_{\text{H}}$.
Future work is needed to fully resolve this issue.
We note that in Ref.~\cite{Breznay2012} disorder effects have been invoked to produce a divergence of $(\sigma_{xx}^{\text{b}})^2$ stronger than that of $\sigma_{xy}^{\text{b}}$,
which leads to a downturn of $\rho_{yx}$ near $T_c$ in amorphous thin films.
Whether this disorder mechanism is relevant in underdoped cuprates is unclear.

Although we have not been able to surmount them, we note that these challenges to transport theory are generic and not restricted to our particular
physical picture.

\subsection{Thermopower} 

We next consider the Seebeck (thermopower) coefficient, $S_{xx}$.  Seebeck data on hole doped cuprates~\cite{Munakata1992,Huang1992,Fujii2002,Badoux2016,CyrChoiniere2017} are no less puzzling than that of the Hall coefficient.
For our focus on low magnetic field and $T>T_c$ in underdoped cuprates, one finds that
$S_{xx}$ shows a broad positive peak at a temperature scale $\sim T^*$ before
it vanishes below $T_c$. 
In contrast, our numerical $S_{xx}$ in Fig.~\ref{fig:Fig3}(c) is almost entirely negative. This is
a consequence of the same band structure needed to explain the Hall coefficient. A small
positive contribution appears at $T$ right above $T_c$
and comes from the bosonic contribution which dominates at $T\approx T_c$: $S_{xx}=\beta_{xx}/\sigma_{xx} \sim \beta_{xx}^{\text{b}}/\sigma_{xx}^{\text{b}}$.
This reflects the fact that $\beta_{xx}^{\text{b}}$ is positive although the fermionic contribution $\beta_{xx}^{\text{f}}$ is negative. 

The issue that the observed sign of $S_{xx}$ is opposite to what one calculates from the simple tight-binding band structure has already been noted in Refs.~\cite{Storey2013,Verret2017}.
In Ref.~\cite{Storey2013}, a pseudogap, whose origin is different from the one we consider here, is used to reconstruct the bare fermionic band structure in order to obtain
a positive $S_{xx}$. However, the theory does not explain the positive $S_{xx}$ at $T>T^*$ where the pseudogap presumably vanishes.
Also considered in the literature was a frequency dependent normal state scattering rate ($\Sigma_0$ in Eq.~\eqref{eq:FermiArcG}) or an anisotropic $\vect{k}$-dependent $\Sigma_0$~\cite{Hussey2008}.
In principle, both of these can lead to a sign change of $S_{xx}$ for $T>T^*$.  Recently a frequency dependence
in $\Sigma_0$ has been assumed to explain the unexpected
sign of $S_{xx}$ in a heavily overdoped cuprate~\cite{Jin2021} where one encounters a similar situation.

Although we have not been able to surmount them, we note, again, that these challenges to transport theory are generic and not restricted to our particular
physical picture.

\subsection{Nernst Coefficient} 

Measurements of the Nernst coefficient $\nu$ on underdoped cuprates~\cite{Xu2000,Wang2001,Wang2006,Chang2011,CyrChoiniere2018} 
have a long history.
The experimental data of $\nu$ at $T>T_c$ in underdoped \LSCO and \BSCCO ~\cite{Xu2000,Wang2001,Wang2006,CyrChoiniere2018} looks qualitatively similar to our results in Fig.~\ref{fig:Fig3}(d).
At $T>T^*$, $\nu$ is small and negative. Below $T^*$ it crosses zero and exhibits a large positive peak centered at a temperature smaller than the magnetic-field dependent $T_c$.
The peak region below $T_c$ is usually attributed to vortex physics which is not included in our theory,
while that above $T_c$ is conventionally attributed to fluctuating Cooper pairs~\cite{Ussishkin2002,Ussishkin2003}.
Early experiments have proposed that the behavior above and below $T_c$ may arise from fluctuating vortices~\cite{Xu2000,Wang2001,Wang2006}, although this scenario has been challenged ~\cite{Behnia2016}.

The Nernst data on underdoped cuprates display some non-universal characteristics.
In contrast to \LSCO and \BSCCO, another pair of cuprates
\YBCO\ and \HgBCO\ exhibit a positive $\nu$ at $T>T^*$~\cite{CyrChoiniere2018},
which is not consistent with the simple fermionic band structure.
In these latter compounds, $\nu$ experiences two sign changes as $T$ drops below $T^*$: it first becomes negative and then becomes
positive again as $T$ approaches $T_c$.
In Ref.~\cite{CyrChoiniere2018}, the first sign change has been taken as evidence against pairing fluctuations playing an important role at $T\approx T^*$, since
the contribution from fluctuating Cooper pairs is always expected to be positive.
However, such a conclusion can be reached only if we better understand the origin of the variation of $\nu$ from one family to another.

In summary, our plots for $\nu$ display reasonably good agreement with experiments  on
\LSCO and \BSCCO~ but are not consistent with the behavior in
\YBCO~ and \HgBCO.
This challenge to the notion of universality is an open question which (to the best of our knowledge)
has not been addressed theoretically.

\subsection{Thermal conductivities: $\kappa_{xx}$ and $\kappa_{xy}$} 

The longitudinal thermal conductivity has a long history in the cuprate field
~\cite{Yu1992,Hirschfeld1996,Sutherland2003}.
The electronic contribution to $\kappa_{xx}$ shows a broad peak as a function of temperature at $T<T_c$, which is believed to arise from an enhancement of the quasiparticle mean free path as superconductivity emerges.
Interestingly, $\kappa_{xx}$ does not exhibit distinctive features as $T$ decreases across $T^*$.
This is in agreement with
the theoretical plot in Fig.~\ref{fig:Fig3}(e) which shows that the bosonic contribution is almost negligible
and the fermionic contribution is very smooth.

Similarly, $\kappa_{xy}$ measured at low fields~\cite{Zeini1999,Krishana1999,Zhang2000,Zhang2001} also shows a broad maximum below $T_c$, which is again attributed to
the increase of the quasiparticle mean free path. Much effort~\cite{Simon1997,Vishwanath2001,Vafek2001,Durst2003,Cvetkovic2015} has been devoted to understanding the behavior of $\kappa_{xy}$ at very low
$T$, deep in the superconducting state, where an intricate interplay
between $d$-wave Bogoliubov quasiparticles and vortices presents a challenge to theory.
While until now there has been relatively little focus on the normal state, a series of
high field $\kappa_{xy}$ measurements~\cite{Grissonnanche2019,Grissonnanche2020}, where the superconductivity is suppressed, were recently performed on underdoped cuprates.
Here it is presumed that this will
reveal the low-$T$ and high field normal state underlying the superconducting phase as well. 
This high-field behavior is beyond the scope of the present paper,
although establishing a baseline
for the behavior even at low fields (with heat magnetization current effects included) is a useful contribution made here.

In Ref.~\cite{Zhang2000}, $\kappa_{xy}$ was measured at low field and over a wide temperature range from $T \gtrsim T_c\sim 90 \mathrm{K}$ all the way to room temperature, which is above
the corresponding $T^* \sim 190 \mathrm{K}$~\cite{LeBoeuf2011}. In this temperature regime, the observed $\kappa_{xy}/B$ is well 
fit by a power law, $1/T^{1.2}$, with no discernible feature at $T^*$. 
The fact that the experimentally observed $\kappa_{xy}$
in Ref.~\cite{Zhang2000} is quite smooth at $T^*$ seems to be consistent with our numerical results in Fig.~\ref{fig:Fig3}(f) where
the bosonic contribution to $\kappa_{xy}$ is relatively small and its divergence near $T_c$ is only logarithmic (see Sec.~\ref{sec:divergence}). (This divergence can be easily cut
off by other effects not included in the current treatment). 
Making a quantitative comparison is not possible at this stage. In particular,
the small contribution from fluctuating Cooper pairs can be easily masked by chiral phonon contributions, if present. 
In this regard, it will be useful in the future to conduct further experiments that connect the low-field high-$T$ regime with the high-field low-$T$ regime.

\section{Quantitative comparison between theoretical and experimental transport coefficients in the cuprates}
\label{app:unit}

The discussions of the previous section already imply that one cannot expect a general agreement between our theory and experiments
over the entire temperature range.
Nevertheless, in Table~\ref{tab:magnitude}, we display our theoretical values of various transport quantities in actual units for two temperatures, $T=1.1 \,  T_c$ and $T=T^*$. 
These values are obtained from Fig.~\ref{fig:Fig3} using the unit conversion from Table.~\ref{tab:unit}. 
Table~\ref{tab:magnitude} does show that
both the sign and magnitudes of the Nernst coefficient $\nu$ at $T=1.1 T_c$ and $T=T^*$ are in good agreement with those of optimally doped \BSCCO~\cite{Wang2006}. 
In this way, a quantitative fit to the longitudinal resistivity seems
to imply a good fit as well to the Nernst coefficient.

\begin{widetext}
\begin{center}
\begin{table*}[h]
\begin{tabular}{| c | c |c | c | c | c | }
\hline \hline
Quantity                    &    \quad  $R_{\text{H}}  [\mathrm{ \frac{cm^3}{C} } ]$  \quad  &   \quad  $S_{xx} [ \mathrm{  \frac{\mu V}{K}  } ]$  \quad    &   \quad $ \nu [ \mathrm{ \frac{\mu V}{K\cdot T}  } ] $   \quad   
&    \quad $ \frac{ \kappa_{xx}}{T}  [ \mathrm{  \frac{m W }{ K^2 \cdot m } } ]$  \quad  &  \quad $ \frac{ \kappa_{xy}}{T B}  [ \mathrm{ \frac{\mu W} { K^2 \cdot m \cdot T }  }]$  \quad   \\
\hline 
$T=1.1 \, T_c$          &     $1.1 \times 10^{-2}$                 &      $-1$                    &        $7\times 10^{-2}$     &     $4.8$     &    $2.0$
\\
[2ex]
\hline
$T=T^*$                    &    $7.0\times 10^{-4}$                  &          $-16$               &        $-5\times 10^{-3}$     &     $3.3$    &   $0.8$
  \\
[1ex]
\hline\hline
\end{tabular}
\caption{Magnitudes of the theoretical transport quantities. While $R_{\text{H}}$ at $T=T^*$ roughly agrees with the corresponding
experimental data for optimally doped \BSCCO~\cite{Konstantinovic2000}, its value at $T=1.1 T_c$ is too big. 
$S_{xx}$ at $T=T^*$ has the right order of magnitude but its sign is opposite to that of underdoped \BSCCO~\cite{Munakata1992}. 
Both the sign and magnitudes of $\nu$ at $T=1.1 T_c$ and $T=T^*$ are in good agreement with those of optimally doped \BSCCO~\cite{Wang2006}. 
The magnitude of $\kappa_{xx}$ at the two temperatures is in rough agreement with experiments~\cite{Yu1992,Hirschfeld1996,Sutherland2003}. The theoretical $\kappa_{xy}/B$ at both temperatures
is an order of magnitude smaller than that observed in Ref.~\cite{Zhang2000}.  }
\label{tab:magnitude}
\end{table*}%
\end{center}

\begin{center}
\begin{table}[htp]
\begin{tabular}{| c | c |c | c | c | c | c | c | c | c | c | }
\hline\hline
Quantity             &  \quad $B$  \quad &  \quad   $\sigma_{xx}$ \quad  &  \quad $ \sigma_{xy}/B$  \quad & \quad $R_{\H}=\frac{\rho_{yx}}{B}$ \quad & \quad $\beta_{xx}$ \quad   &   \quad $ \beta_{xy}/B$ \quad   &   $S_{xx} = \frac{\beta_{xx}}{\sigma_{xx}}$    &  $ \nu $     &   $\kappa_{xx}/T$   &  $\kappa_{xy}/T/B$    \\
\hline
Latt. unit   &  $\frac{\hbar}{e a^2} $   & $\frac{e^2}{\hbar d}$   &   $\frac{e^3 a^2}{\hbar^2 d}$   &  $\frac{a^2 d}{|e|}$ & $\frac{e k_B }{\hbar d}$  &  $\frac{e^2 k_B  a^2}{\hbar^2 d}$   &   $ \frac{k_B}{e} $   & 
$\frac{a^2 k_B}{\hbar}$  &  $ \frac{ k_B^2}{\hbar d}$  & $\frac{e k_B^2 a^2}{\hbar^2 d}$   \\
[2ex]
\hline
Numerical value &  $4.56\times 10^3$ & $3.17$ & $6.96 \times 10^{-4}$ & $6.91 \times 10^{-4}$ & $27.3$   &  $6.0\times 10^{-3}$ & $86.2$  & $1.89\times 10^{-2}$ & $2.36$ & $0.52$  \\
[1ex]
\hline 
Units &  $\mathrm{Tesla \; (T) }$ &  $ \mathrm{ \frac{1}{m\Omega\cdot cm} } $ &  $\mathrm{ \frac{1}{m\Omega\cdot cm\cdot T} }$  &  $ \mathrm{ \frac{cm^3}{C}  }$  & $ \mathrm{ \frac{A}{K\cdot m} }$  
& $ \mathrm{  \frac{A}{K\cdot m \cdot T}  } $   & $ \mathrm{ \frac{ \mu V}{ K} } $   & 
$\mathrm{ \frac{ \mu V}{ K\cdot T } } $  &  $  \mathrm{  \frac{mW}{K^2 \cdot m }  } $  &  $ \mathrm{   \frac{\mu W}{K^2 \cdot m \cdot T }  }$   \\
[2ex]
\hline\hline
\end{tabular}
\caption{Units used for different quantities. To obtain the second and third rows we have used $a=3.8$\AA, $d=30.7/4=7.67$ \AA~for cuprates~\cite{Petricek1990}, where $a$ is the in-plane
lattice constant and $d$ is the inter-layer
spacing per CuO$_2$ plane.
Although all quantities are calculated for 2d in the main text, we switch to 3d in this table and show explicitly their dependence on the third dimensional length scale, $d$,
for easier comparison to experiments. }
\label{tab:unit}
\end{table}%
\end{center}
\end{widetext}

\newpage
\bibliography{References}   

\begin{thebibliography}{127}
\bibitem [{BCS()}]{BCSnote}%
  \BibitemOpen
  \href@noop {} {}\bibinfo {note} {We argue that, above $T_c$, one should be
  cautious when comparing the oversimplified BCS form of the Bogoliubov
  quasiparticle dispersion with that extracted from the ARPES energy
  distribution curves, in the absence of clear coherence peaks and given the
  expected large spectral broadening due to both high $T$ and incoherent
  pairing processes.}\BibitemShut {Stop}%
\end{thebibliography}

\end{document}